\documentclass[review,3p]{elsarticle}
\biboptions{sort&compress}
\usepackage[utf8]{inputenc}
\usepackage{caption}
\usepackage{enumitem}
\usepackage{amssymb}
\usepackage{amsmath}
\usepackage{epstopdf}
\usepackage{subeqnarray}
\usepackage{fancyhdr}
\usepackage{accents}
\usepackage{graphicx}

\usepackage{amsmath,amssymb,bm,bbm}
\usepackage[colorlinks=true,
linkcolor=blue,
urlcolor=blue,
citecolor=gray] {hyperref}
\usepackage[all]{hypcap}
\usepackage{ulem}

\usepackage{subcaption}
\usepackage{epstopdf}
\usepackage{booktabs}
\usepackage{relsize}
 \usepackage{lineno}

\journal{}

\begin{document}
\begin{frontmatter}

\title{Combination of operational modal analysis algorithms to identify modal parameters of an actual centrifugal compressor}

\author[UFRJ,UFRJ2]{Leandro O. Zague}
\ead{zague@etm-turbo.com}

\author[UFRJ]{Daniel A. Castello \corref{cor1}}
\ead{castello@mecanica.coppe.ufrj.br}
		
\author[CEPEL]{Carlos F.T. Matt }
\ead{cfmatt@cepel.br}

\address[UFRJ]{Federal University of Rio de Janeiro, Department of Mechanical Engineering, Rio de Janeiro, Brazil}
\address[UFRJ2]{ETM Energy \& Turbomachinery, Rio de Janeiro, Brazil}
\address[CEPEL]{Laboratory of Mechatronics and Structural Dynamics, Department of Materials and Mechatronics, CEPEL, Rio de Janeiro, Brazil}

\cortext[cor1]{Corresponding author: Daniel Alves Castello . 
E-mail: castello@mecanica.coppe.ufrj.br, Phone:+55 21 3938-8385.}

\begin{abstract}
The novelty of the current work is precisely to propose a statistical procedure to combine estimates of the modal parameters provided by any set of Operational Modal Analysis (OMA) algorithms so as to avoid preference for a particular  one and also to derive an approximate joint probability distribution of the modal parameters, from which engineering statistics of interest such as mean value and variance are readily provided. The effectiveness of the proposed strategy is assessed considering measured data from an actual centrifugal compressor. The statistics obtained for both forward and backward modal parameters are finally compared against modal parameters identified during standard stability verification testing (SVT) of centrifugal compressors prior to shipment, using classical Experimental Modal Analysis (EMA) algorithms. The current work demonstrates that combination of OMA algorithms can provide quite accurate estimates for both the modal parameters and the associated uncertainties with low computational costs.
\end{abstract}
		
\end{frontmatter}
	
%\linenumbers
	
%% main text

%%% END OF ORIGINAL INTRODUCTION
\section{Introduction}

Rotating machinery equipment is present in almost any industrial facilities/plants. The development of strategies to assess its operational health is of great interest to academia and industrial sectors inasmuch its effectiveness may lead to more optimized operational scenarios. These strategies in principle may be built based on different types of measured data such as the modal ones, for example. Nevertheless, measurements of input excitation are rarely available in industrial sectors and if one is interested in using modal data it is required to use estimators solely based on ambient vibration data such as Operational Modal Analysis (OMA) algorithms. In this regard, the current work brings a proposal to combine OMA estimates provided by a set of algorithms based on a statistical framework.  One aspect to be highlighted is that the proposal of combination here completely avoids  any attempt to compare OMA algorithms inasmuch as model order selection, spurious modes rejection and harmonic rejection, to cite a few, are issues that make such comparisons  practically impossible.

More specifically, the novelty of the current work is precisely to propose a statistical procedure to combine estimates of the modal parameters provided by any set of OMA algorithms so as to avoid preference for a particular OMA algorithm (and thus biasing the end results) and also to derive an approximate Gaussian joint probability distribution of the modal parameters, from which engineering statistics of interest such as mean value and variance are readily provided. Since the excitation levels in OMA are generally quite low, recent works have focused on the uncertainty quantification of the modal parameters  provided by OMA algorithms \cite{Sedehietal2020,Auetal2018}. The proposed statistical strategy also provides an uncertainty quantification for the estimated modal parameters but avoids full Bayesian formalism and computation adopted in references \cite{Sedehietal2020,Auetal2018}, which improves its computational performance.

Regarding the assessment of the proposed approach, it is applied to experimental data collected for an actual nine-impeller centrifugal compressor. The statistics obtained for both forward and backward modal parameters are finally compared against modal parameters identified during standard stability verification testing (SVT) of centrifugal compressors {\it{prior to shipment}}, using classical EMA algorithms. The comparison reveals that the proposed strategy provides quite accurate results for the modal parameters and also estimates for the associated uncertainty in a computationally-efficient manner. More importantly, it takes into account not only measurement uncertainties but also different levels of model uncertainties (induced by modeling errors) inherent to each OMA identification algorithm. Finally, one aspect to be highlighted is that the application of OMA algorithms to identify modal parameters of centrifugal compressors is far from being completely understood by the rotating machinery community. To the authors' best knowledge, relatively few works are devoted to the application of OMA algorithms to centrifugal compressors; the recent work of Salehi {\it et al}.\ \cite{Salehietal2018} is the first published work dealing with application of OMA algorithms to identify the modal parameters of an actual centrifugal compressor. 

We finish by remarking that the proposed strategy may provide fast and accurate estimates of the modal parameters along with the associated uncertainties, which may improve not only future rotordynamic designs but also the predictive accuracy of existing computational models and real-time assessment of rotor's stability condition. The manuscript is organized as follows. Section 2 presents a brief literature review. Section 3 briefly summarizes the theoretical fundamentals of OMA algorithms. Section 4 describes the proposed statistical strategy to combine estimates of modal parameters identified with different OMA algorithms. Section 5 describes the %two dynamical systems investigated: (i) a computational finite-element model of a rotor-bearing system and (ii) an 
actual centrifugal compressor investigated. Results obtained for the modal parameters and associated uncertainties are reported and discussed in Section 6. Final remarks and perspectives for future works are highlighted in Section 7.

\section{Brief Literature Review}

OMA algorithms are alternatives to the classical experimental modal analysis (EMA) algorithms to identify the modal parameters using solely output measured data. Particularly for centrifugal compressors, stability tests become a simpler task and OMA algorithms ease the assessment of compressor's stability during its field operation, as coupling of magnetic exciters to provide non-synchronous excitation is avoided. These features make OMA a very attractive vibration troubleshooting tool. Although largely employed for modal identification of large-scale civil engineering infrastructures (e.g., tall buildings, bridges, stadiums and oil rigs) \cite{Wagneretal2022,JVC2}, OMA is indeed not well-acknowledged by the rotating machinery community.

The theory of OMA algorithms is developed based on the assumption that the excitation, even though unknown, may be decomposed into stochastic and deterministic components. Particularly for centrifugal compressors, the stochastic component arises from turbulent gas flow through internal channels of the compressor and the deterministic component arises from unavoidable unbalance in such large machines. The combined unknown excitation is thus modelled by a filtered stationary, zero
mean, Gaussian white noise excitation as described by Brincker and Ventura \cite{OMABrincker}. For a successful analysis, the stochastic component of the excitation must have enough bandwidth to cover the frequency range of interest and must be strong enough to provide a good signal-to-noise ratio. The lack of control over the stochastic excitation is a disadvantage to be addressed by OMA algorithms and claimed by many authors as responsible for larger uncertainties induced in the modal parameters when compared with EMA algorithms. There are many applications of OMA in mechanical engineering reported in the literature: for on-the-road modal analysis of cars; in-flight modal analysis of airplanes; modal testing of spacecraft during launch; modal testing of engines during startup and shutdown; modal testing of small-scale wind turbines in laboratory \cite{Wagneretal2022}, to name just a few. For civil engineering applications, OMA methods have become the primary tool for modal testing and the number of reported case studies is abundant \cite{reynders2012system}. On the other hand, the successful application of OMA methods to actual centrifugal compressors is far less documented.

OMA algorithms are classified as time-domain and frequency-domain methods. Akaike \cite{akaike1974markovian} and Bendat and Piersol \cite{bendat1980engineering} were pioneers in recognizing that impulse responses and free decays can be obtained from the output (or response) correlation functions, when a white noise excitation is applied. Afterwards, other time-domain methods were developed such as the Ibrahim time domain \cite{ibraham1977method}, the polyreference method \cite{vold1982multi}, the eigensystem realization algorithm (ERA) \cite{Pappa}, the natural excitation technique \cite{NExT} and the stochastic subspace identification (SSI) methods \cite{van2012subspace}, to name a few. In 2000's, efforts were concentrated on uncertainty quantification of modal parameters identified from OMA algorithms. Full Bayesian approach for time-domain OMA was developed by Yuen and Katafygiotis \cite{yuen2001bayesian}. Later, Yuen and Katafygiotis \cite{YuenKatafygiotis2003} extended their previous time-domain Bayesian method to identify modal parameters using as observed data the `raw' (i.e., no filtering, no windowing) fast Fourier Transform of time-domain responses within selected frequency bands enclosing eigenfrequencies, yielding the fast FFT Bayesian OMA (coined as fast BAYOMA by those authors). Nevertheless, a recent review in Yuen and Kuok \cite{yuen2011bayesian} has indicated potential significant bias for BAYOMA. Reynders {\it et al.}\ \cite{reynders2008uncertainty} have proposed a method to estimate the variance of modal parameters extracted from a single ambient vibration test with SSI methods. Meanwhile, efforts were also concentrated on harmonic signal elimination, as harmonics coinciding with natural frequencies can bring challenges to OMA algorithms; see, for instance, Refs.\ \cite{modak2010harmonics,peeters2007removing,qi2008vibration}. Nowadays, efforts in the aforementioned two issues are still ongoing. Recently, a new orthogonal projection-based method was proposed in Gres {\it et al.} \cite{gres2019orthogonal} for harmonic signal removal. Reynders {\it et al.} \cite{reynders2016uncertainty} extended their previous work to validate estimates for the variance of modal parameters obtained from a single ambient vibration test using SSI methods. Those authors also validated the maximum likelihood approach proposed by Pintelon {\it et al.}\  \cite{pintelon2007uncertainty} for uncertainty quantification of the identified modal parameters. Au {\it et al.}\  \cite{Auetal2018} revisited the fast BAYOMA proposed by Yuen and Katafygiotis \cite{YuenKatafygiotis2003} and, by using first-order asymptotic analysis, they derived analytical expressions for the leading (zeroth) and first-order expansion coefficients of the posterior C.o.V (standard-deviation/mean) of each identified modal parameter (eigenfrequency, damping ratio and mode shape) in a context of well-separated eigenfrequencies, small damping and very large data length. These expressions are now referred to as the `uncertainty laws'. Those authors have demonstrated that the leading (zeroth) order expansion coefficients give the achievable limit of BAYOMA precision when a parameter coined as {\it modal signal-to-noise ratio} is infinite.  More recently, Sedehi {\it et al.}\ \cite{Sedehietal2020} have extended BAYOMA by incorporating a hierarchical formulation to rigorously account for both the variability among data uncertainty (when using multiple data sets) and the identification precision for each data uncertainty. The motivation for the hierarchical BAYOMA formulation proposed by Sedehi {\it et al.}\ \cite{Sedehietal2020} has been the criticism about earlier Bayesian formulations, as observations had shown that the variability on identified modal parameters over different experiments (data sets) significantly exceeds the Bayesian estimation of uncertainty. Notwithstanding the foregoing discussions, the authors have resorted to simplifying assumptions as well-separated eigenfrequencies and small damping in order to alleviate the huge computational efforts associated with BAYOMA. These assumptions are not always verified for rotating machinery applications and certainly would also introduce modeling errors; such modeling errors have not been accounted for in the aforementioned hierarchical Bayesian formulations.

Particular applications of OMA to centrifugal compressors are reported in Salehi {\it et al}.\ \cite{Salehietal2018}, Guglielmo {\it et al.} \cite{Guglielmo2}, Guglielmo {\it et al.} \cite{Guglielmo15} and Carden {\it et al}.\ \cite{Loyds}. In Salehi {\it et al}.\ \cite{Salehietal2018} modal parameters of a four-stage centrifugal compressor were extracted and compared using three OMA algorithms, namely, the frequency domain decomposition (FDD), the enhanced frequency domain decomposition (EFDD) and the stochastic subspace identification (SSI). Those authors concluded that the SSI algorithm had higher accuracy than the FDD and EFDD algorithms; nevertheless, FDD provided better results when system damping was low in one of the identified modes. In Guglielmo {\it et al.} \cite{Guglielmo2}, OMA and EMA were applied to a low pressure LNG compressor at different loading conditions during a full-load stability test. In Guglielmo {\it et al.} \cite{Guglielmo15}, OMA and EMA were applied to a high-pressure reinjection compressor and also to a rotordynamic model. Those authors highlighted aspects of OMA specific to centrifugal compressors and compared methods in the two groups for several damping levels. In Carden {\it et al.}\ \cite{Loyds}, OMA was used to confirm that the first forward mode of a 500 kW compressor was actually stable and measurements corroborated the predicted behavior. A particular OMA algorithm (multiple-output backward auto-regressive models or MOBAR for short) combined with blocking test excitation was successfully applied to extract modal parameters in a laboratory-scale rotor in Cloud \cite{Cloud2007} and Cloud {\it et al.} \cite{Hunter_damp}, in cage induction motors in Holopainen {\it et al.} \cite{Holopainen} and also in multistage centrifugal compressors during factory stability tests in Pettinato {\it et al.} \cite{Hunter} and Noronha {\it et al.} \cite{Noronha}. Most stability tests on centrifugal compressors are performed with EMA algorithms and successful applications of OMA to actual centrifugal compressors are far less reported in the scientific literature. 

\section{Theoretical Fundamentals of OMA Algorithms}\label{sec:oma_algorithms}

This section summarizes the theoretical basis underpinning OMA algorithms.  Let ${\bf y}(t)$ denotes the column vector containing the measured data set observed from a  dynamical system. The output correlation matrix ${\bf R}_{\bf y}$ is defined as the ensemble average 
\begin{equation}
{\bf R}_{\bf y}(\tau) \equiv \mathbb{E}[{\bf y}(t) {\bf y}^{\rm T}(t+\tau)] \label{outputcovariancematrixcontinuoussignals}
\end{equation}
and, for stationary random processes as ${\bf y}(t)$, it depends only on the time lag $\tau$. All the response signals analyzed herein have its mean value subtracted from the raw signal. Now, by using the convolution relationship between the input ${\bf u}(t)$ and output ${\bf y}(t)$ for a multiple-input multiple-output (MIMO) system, one may rewrite Eq.\ (\ref{outputcovariancematrixcontinuoussignals}) as

\begin{equation}\label{eq:fundtheo}
{\bf R}_{{\bf y}}(\tau)={\bf H}(-\tau)*{\bf R}_{\bf u}(\tau)*{\bf H}^T(\tau)
\end{equation}
where ${\bf H}(\tau)$ is the impulse response function matrix (IRM), ${\bf R}_{{\bf u}}(\tau)$ and ${\bf R}_{{\bf y}}(\tau)$ denote, respectively, the correlation matrices of the input and output, and the symbol $*$ denotes a convolution integral. The input and output correlation matrices are related to each other through the impulse response function matrix ${\bf H}(\tau)$ via a double convolution. 

By taking the Fourier transform $\mathcal{F}\{\bullet \}$ on both sides of Eq.\ (\ref{eq:fundtheo}) one obtains a relationship between the input and output power spectral density matrices, respectively denoted by ${\bf G}_{\bf u}(\Omega)$ and ${\bf G}_{\bf y}(\Omega)$, as follows
\begin{equation}\label{eq:fundtheo2}
{\bf G}_{\bf y}(\Omega) = {\bf {\tilde{H}}}(-i\Omega) \, {\bf G}_{\bf u}(\Omega) \, {\bf {\tilde{H}}}^T(i\Omega)
\end{equation}
\noindent where $\mathbf{G}_{\mathbf{y}}(\Omega) = \mathcal{F}\{\mathbf{R}_{\mathbf{y}}(t)\}$  and $\mathbf{G}_{\mathbf{u}}(\Omega) = \mathcal{F}\{\mathbf{R}_{\mathbf{u}}(t)\}$ and $\mathbf{\tilde{H}}(i\Omega)$ corresponds to frequency response function (FRF) matrix. Equation (\ref{eq:fundtheo2}) represents the fundamental theorem of stochastic vibrations for MIMO systems in the frequency domain. Time-domain OMA methods use as input data the output correlation matrix $\mathbf{R}_{\mathbf{y}}(\tau)$ whereas their frequency-domain counterparts use as input data the output power spectral density matrix ${\bf G}_{\bf y}(\Omega)$. In the next subsections, one briefly describes three widely-used time-domain OMA methods that will be used in this work: the covariance-driven SSI \cite{JVC1}, data-driven SSI and MOBAR. Although mathematical details may be encountered elsewhere \cite{OMABrincker,van2012subspace}, we decided to report here the main equations for easy reference and not to compromise the understanding by readers not well-familiarized with OMA algorithms.

\subsection{Stochastic Subspace Identification}\label{SSI_algorithms}

The stochastic subspace identification (SSI) method aims to identify a stochastic state space model for a vibrating dynamical system  \cite{van2012subspace} such the one shown in Eq.(\ref{statespace2})  \cite{PeetersDeRoeck1999}:
\begin{eqnarray}
{\bf x}_{k+1} &=& {\bf A} \, {\bf x}_{k} + {\bf w}_{k} \label{statespace1}\\
{\bf y}_{k} &=& {\bf C} \, {\bf x}_{k} + {\bf v}_{k} \label{statespace2}
\end{eqnarray}
\noindent where the excitation is assumed to be a white noise,  ${\bf y}_{k} \in \mathbb{R}^{\ell \times 1}$ denotes the measurements of the $\ell$ outputs at discrete time instant $k$ (i.e., ${\bf y}_{k} = {\bf y}(t_{k})$, with $t_{k} = k \, \Delta t$, $\Delta t$ being the sampling time and $k \in \mathbb{N}$); ${\bf x}_{k} \in \mathbb{R}^{n \times 1}$ is the state vector and $n$ denotes the system order. The vector ${\bf w}_{k} \in \mathbb{R}^{n \times 1}$ is the process noise due to disturbances and modelling inaccuracies and also models the white noise input and ${\bf v}_{k} \in \mathbb{R}^{\ell \times 1}$ is the measurement noise. Process and measurement noises are both unmeasurable, but {\it{assumed}} to be zero mean, white and with covariance matrices such that
\begin{equation}
\mathbb{E}\left[\left(
\begin{array}{c}
{\bf{w}}_{p} \\
{\bf{v}}_{p}
\end{array}
\right) \left(
\begin{array}{cc}
{\bf w}_{q}^{\rm T} & {\bf v}_{q}^{\rm T}
\end{array} \right)\right] = \left[
\begin{array}{cc}
{\bf Q} & {\bf S} \\
{\bf S}^{\rm T} & {\bf R}  
\end{array} \right] \delta_{pq}
\end{equation}
where $\delta_{pq}$ denotes the Kronecker delta and $p, q$ are two arbitrary time instants. The main goal of SSI methods is to estimate the system order $n$, and discrete state and output matrices ${\bf A} \in \mathbb{R}^{n \times n}$ and ${\bf C} \in \mathbb{R}^{\ell \times n}$ from which one may compute the eigenfrequencies, damping ratios and mode shapes of the vibrating structure. There are two main SSI methods, namely the covariance-driven and the data-driven SSI.

\paragraph{Covariance-driven SSI (COV-SSI)} The covariance-driven SSI algorithm uses direct estimates of the output covariances to identify the modal parameters. The output measurements are gathered in a block Hankel matrix ${\mathbb{H}} \in \mathbb{R}^{2 \ell i \times N}$. The Hankel matrix can be divided into a past and a future part, as follows
\begin{equation} \label{eq:Hankel1}
{\mathbb{H}} =  
\begin{bmatrix}
{\bf Y}_{0|i-1} \\ \cline{1-1}
{\bf Y}_{i|2i-1}
\end{bmatrix} =
\begin{bmatrix}
{\bf Y}_{p} \\ \cline{1-1}
{\bf Y}_{f}
\end{bmatrix}
\end{equation}
where ${\bf Y}_{p} = {\bf Y}_{0|i-1} \in \mathbb{R}^{\ell i \times N}$ and ${\bf Y}_{f} = {\bf Y}_{i|2i-1} \in \mathbb{R}^{\ell i \times N}$ are, respectively, the past and future measurements. Regarding the components of ${\mathbb{H}}$, interested readers may find it in Appendix Eq.(\ref{eq:Hankel1Ap}). %The output data need to be scaled by a factor of $1/\sqrt{N}$ to be consistent with the definition of the output covariances, as shortly demonstrated.

The first step of the COV-SSI algorithm is to construct a block-Toeplitz matrix ${\bf T}_{1|i} \in \mathbb{R}^{\ell i \times \ell i}$ whose components are the output covariance matrices for different time lags, as can be seen in Eq.\ (\ref{eq:Toeptitz1}):
\begin{equation} \label{eq:Toeptitz1}
{\bf T}_{1|i} = {\bf Y}_{f} \, {\bf Y}_{p}^{T} 
\end{equation}
Detailed information about the  components of the matrix ${\bf T}_{1|i}$ may be found in Appendix Eq.(\ref{eq:Toeptitz1Ap}).
The block Toeplitz matrix ${\bf T}_{1|i}$ can be decomposed into the {\it extended observability matrix} ${\boldsymbol \Gamma}_{i} \in \mathbb{R}^{\ell i \times n}$ and the {\it reversed extended stochastic controllability matrix} ${\boldsymbol \Delta}_{i} \in \mathbb{R}^{n \times \ell i}$ as follows
\begin{equation} %\label{eq:Toeplitz2}
{\bf T}_{1|i}=
 {\boldsymbol \Gamma}_{i} \, {\boldsymbol \Delta}_{i}
\label{blockHankelmatrixdecomp}
\end{equation}
where ${\boldsymbol \Gamma}_{i} \equiv [{\bf C} \,\, {\bf C} {\bf A} \,\, {\bf C} {\bf A}^{2} \,\, \dots \,\, {\bf C} {\bf A}^{i-1}]^{T}$ and ${\boldsymbol \Delta}_{i} \equiv [{\bf A}^{i-1} {\bf G} \,\, \dots \,\, {\bf A}^{2} {\bf G} \,\, {\bf A} {\bf G} \,\, {\bf G}]$ and detailed information about the structure of its components is found in Appendix Eq.(\ref{blockHankelmatrixdecompAp}). The second step is to perform a singular value decomposition (SVD) of the block Toeplitz matrix ${\bf T}_{1|i}$ to obtain

\begin{eqnarray}
{\bf T}_{1|i} &=& {\bf U} \, {\bar {\bf S}} \, {\bf V}^{T} = [{\bf U}_{1} \,\, {\bf U}_{2}] \, 
\begin{bmatrix}
{\bf S}_{1} & {\bf 0} \\
{\bf 0} & {\bf 0}
\end{bmatrix}
\begin{bmatrix}
{\bf V}_{1}^{T} \\
{\bf V}_{2}^{T} 
\end{bmatrix} = {\bf U}_{1} \, {\bf S}_{1} \, {\bf V}_{1}^{T} = ({\bf U}_{1} \, {\bf S}_{1}^{1/2}) ({\bf S}_{1}^{1/2} \, {\bf V}_{1}^{T})
\label{blockHankelmatrixSVD}
\end{eqnarray}

\noindent where ${\bf U} \in \mathbb{R}^{\ell i \times \ell i}$ and ${\bf V} \in \mathbb{R}^{\ell i \times \ell i}$ are both orthogonal matrices; ${\bar {\bf S}} \in \mathbb{R}^{\ell i \times \ell i}$ is a diagonal matrix containing all the singular values in descending order and ${\bf {\bar S}}_{1} \in \mathbb{R}^{n \times n}$ is a diagonal matrix containing only the non-zero singular values in descending order. Comparing Eqs.\ (\ref{blockHankelmatrixdecomp}) and (\ref{blockHankelmatrixSVD}) one immediately obtains ${\boldsymbol \Gamma}_{i} = {\bf U}_{1} \, {\bf S}_{1}^{1/2}$ and ${\boldsymbol \Delta}_{i} = {\bf S}_{1}^{1/2} \, {\bf V}_{1}^{T} $.
Regarding the system order $n$, here we adopted a simpler and widely-used criteria to determine it gradually increasing the dimension of the matrix ${\bf S}_{1}$ (including more nonzero singular values) and checking for stabilization of the identified eigenfrequencies and damping ratios. As for the system matrices, the first block row matrix of  ${\boldsymbol \Gamma}_{i}$ and the last block column matrix of ${\boldsymbol \Delta}_{i}$ are, respectively, the output matrix ${\bf C}$ and the ``next state-output" matrix ${\bf G}$. The system matrix ${\bf A}$ is computed using a block Toeplitz matrix ${\bf T}_{2|i+1}$ defined as
\begin{eqnarray} 
{\bf T}_{2|i+1} &=& {\boldsymbol \Gamma}_{i} \, {\bf A} \, {\boldsymbol \Delta}_{i}
\label{blockHankelmatrix2decomp}
\end{eqnarray}
\noindent where detailed information about its components may be found in Appendix Eq.(\ref{blockHankelmatrix2decompAp}). 

By substituting Eq.\ ${\boldsymbol \Gamma}_{i} = {\bf U}_{1} \, {\bf S}_{1}^{1/2}$ and ${\boldsymbol \Delta}_{i} = {\bf S}_{1}^{1/2} \, {\bf V}_{1}^{T} $ into the last equality in Eq.\ (\ref{blockHankelmatrix2decomp}) yields
\begin{equation}
{\bf A} = {\bf S}_{1}^{-1/2} \, {\bf U}_{1}^{T} \, {\bf T}_{2|i+1} \, {\bf V}_{1} \, {\bf S}_{1}^{-1/2} 
\end{equation}
The final step of the algorithm is to perform an eigenvalue decomposition of the system matrix ${\bf A}$, i.e., ${\bf A} = {\bf \Phi} \, {\bf \Lambda} \, {\bf \Phi}^{-1}$, by computing the diagonal matrix ${\bf \Lambda} = [{\rm diag}(\lambda_{j})] \in \mathbb{C}^{n \times n}$, $j = 1, 2, \dots, n$, containing the eigenvalues $\lambda_{j}$ of the matrix $\mathbf{A}$ and the orthogonal matrix ${\bf \Phi} \in \mathbb{C}^{n \times n}$ whose $j^{th}$ column is the eigenvector of ${\bf A}$ associated with eigenvalue $\lambda_{j}$. The eigenvalues of the corresponding continuous state-space model, $\lambda_{cj}$, are readily computed as $\lambda_{cj} = \frac{\ln{\lambda_j}}{\Delta t}$ and they appear as complex conjugated pairs. Hence, the eigenfrequencies $\omega_{j}$ and damping ratios $\zeta_{j}$, $j = 1, 2, \dots, n$, are finally computed as
\begin{equation}
\omega_{j} = \sqrt{{\rm Re}^{2}(\lambda_{cj}) + {\rm Im}^{2}(\lambda_{cj})} \;\;\; {\rm and} \;\;\; \zeta_{j} = - \frac{{\rm Re}(\lambda_{cj})}{\omega_{j}}   
\end{equation}
and the mode shapes ${\bf V} \in \mathbb{C}^{\ell \times n}$ are found as ${\bf V} = {\bf C} \, {\bf \Phi}$. 
%%%%%%%%%%%%%%%%%%%%%%%%%%%%%%%%%%%%%%%%%%%%%%
\paragraph{Data-driven SSI} The data-driven SSI algorithm starts with the previously-defined block Hankel matrix ${\mathbb{H}} \in \mathbb{R}^{2 \ell i \times N}$ separated in two data sets: (i) the output responses in past times ${\bf Y}_{p} \equiv {\bf Y}_{0|i-1}$ and (ii) the output responses in future times ${\bf Y}_{f} \equiv {\bf Y}_{i|2i - 1}$ as shown in Eq.\ (\ref{eq:Hankel1}). The algorithm proceeds with projecting the row space of the future outputs into the row space of the past outputs which is defined as the expected values of future outputs, conditioned on the known past outputs \cite{Andersen,PeetersDeRoeck2000}, i.e.,
\begin{equation}
{\bf {\cal P}_{i}} \equiv {\mathbb{E}}({\bf Y}_{f}| {\bf Y}_{p}) = {\bf Y}_{f} \, {\bf Y}_{p}^{T} \left({\bf Y}_{p} \, {\bf Y}_{p}^{T}\right)^{-1} \, {\bf Y}_{p} 
\label{eq:SSIO}
\end{equation}
%Notice that the first two terms in Eq.\ (\ref{eq:SSIO}) are equivalent to the block Toeplitz matrix with covariances used in the COV-SSI algorithm, further decomposed into an observability and a controllability matrix. According to Reynders \cite{reynders2012system} and Peeters and De Roeck \cite{PeetersDeRoeck2001}, the notions of covariances and projections are closely-related because both are aimed to cancel out the uncorrelated noise. The next term in Eq.\ (\ref{eq:SSIO}) is the inverse of a block Toeplitz matrix constructed from the past output correlation matrices. 
The main theorem of stochastic subspace identification states that the projection matrix ${\bf {\cal P}}_{i}$ can be factorized as the product of the previously-defined observability matrix ${\bf \Gamma}_{i}$ and another matrix ${\hat {\bf X}}_{i}$, referred to as the Kalman filter state matrix
\begin{equation}
{\bf {\cal P}}_{i} = {\bf \Gamma}_{i} \,\left[
\begin{array}{ccccc}
{\hat {\bf x}}_{i} & {\hat {\bf x}}_{i+1} & {\hat {\bf x}}_{i+2} & \dots & {\hat {\bf x}}_{i+N-1}  
\end{array} \right] = {\bf \Gamma}_{i} \, {\hat {\bf X}}_{i}
\label{projectionmatrix}
\end{equation}
where ${\hat {\bf x}}_{i}$ denotes the optimal Kalman filter prediction for the state vector ${\bf x}_{i}$ by making use of the output measurements up to the time instant $i$ (${\bf Y}_{0|i}$) and the available system matrices and noise covariances. By comparing Eqs.\ (\ref{eq:SSIO}) and (\ref{projectionmatrix}), one may notice that ${\hat {\bf X}}_{i} = {\bf \Delta}_{i} \, \left({\bf Y}_{p} \, {\bf Y}_{p}^{T}\right)^{-1} \, {\bf Y}_{p}$, where ${\bf \Delta}_{i}$ is the previously-defined controllability matrix. %The columns of the projection matrix may be interpreted as free decays given by different initial conditions specified by the Kalman states \cite{Andersen}. 

The next step of the DATA-SSI algorithm is to left multiply the projection matrix ${\bf {\cal P}}_{i}$ by a weighting matrix ${\bf W}_{1}$ and to right multiply ${\bf {\cal P}}_{i}$ by another weighting matrix ${\bf W}_{2}$. Three different definitions are used for these weighting matrices, leading to three different DATA-SSI algorithms: unweighted principal component (UPC), principal component (PC) and canonical variate analysis (CVA). Table \ref{tab:algos} presents ${\bf W}_{1}$ and ${\bf W}_{2}$ for each of the three DATA-SSI algorithms where ${\bf I}_{1} \in \mathbb{R}^{\ell i \times \ell i}$ and ${\bf I}_{2} \in \mathbb{R}^{N \times N}$ denote identity matrices.
\begin{table}[h!]
\caption{Definition of weighting matrices ${\bf W}_{1}$ and ${\bf W}_{2}$ for the three DATA-SSI algorithms implemented}
	\label{tab:algos}
	\centering
	{\footnotesize
		\begin{tabular}{|c|c|c|}
			\hline
			\verb|| & ${\bf W}_{1}$ & ${\bf W}_{2}$\\
			\hline
			PC &  ${\bf I}_{1}$ &  ${\bf Y}_{p}^{T} \, (\frac{1}{N} {\bf Y}_{p} \, {\bf Y}_{p}^T)^{-1/2} \, {\bf Y}_{p}$  \\
			\hline
			UPC &  ${\bf I}_{1}$ &  ${\bf I}_{2}$ \\
			\hline
			CVA &  $(\frac{1}{N} {\bf Y}_{f} \, {\bf Y}_{f}^{T})^{-1/2}$ &  ${\bf I}_{2}$ \\
			\hline
	\end{tabular}}
\end{table}
The final step of the algorithm is to perform a (truncated) singular value decomposition of the weighted projection matrix, i.e.,
\begin{equation}
{\bf W}_{1} \, {\bf {\cal P}}_{i} \, {\bf W}_{2} = {\bf U} \, {\bf S} \, {\bf V}^{T} = 
\left[\begin{array}{cc}
{\bf U}_{1} & {\bf U}_{2}  \end{array} \right] 
\left[\begin{array}{cc}
{\bf S}_{1} & {\bf 0} \\
{\bf 0} & {\bf 0} 
\end{array} \right] 
\left[
\begin{array}{c}
{\bf V}_{1}^{T} \\
{\bf V}_{2}^{T} 
\end{array} \right] = {\bf U}_{1} \, {\bf S}_{1} \, {\bf V}_{1}^{T}
\end{equation}
hence, ${\bf \Gamma}_{i} = {\bf U}_{1} \, {\bf S}_{1}^{1/2}$ and ${\hat {\bf X}}_{i} = {\bf S}_{1}^{1/2} \, {\bf V}_{1}^{T}$. % Regarding the system order, in practice it can be estimated by gradually increasing the number of near-zero singular values. A common practice is to increase the system order and extract the modal parameters by creating a stabilization diagram from which one may distinguish between physical and numerical modes.  

The output matrix ${\bf C}$ may be found from the first block row matrix of ${\bf \Gamma}_{i}$, as before. The system matrix ${\bf A}$ is computed using another projection matrix ${\bf {\cal P}}_{i-1}$, defined by shifting one block row down the separation between past and future outputs in Eq.\ (\ref{eq:Hankel1}), i.e., ${\bf {\cal P}}_{i-1} \equiv \mathbb{E}({\bf Y}_{i+1|2i-1}|{\bf Y}_{0|i}) = {\bf \Gamma}_{i-1} \, {\hat {\bf X}}_{i+1}$, with ${\bf \Gamma}_{i-1}$ directly obtained from ${\bf \Gamma}_{i}$ after deleting its last $\ell$ rows. The shifted state vector ${\hat {\bf X}}_{i+1}$ is obtained after truncated singular value decomposition of (weighted) one block-shift-down projection matrix ${\bf {\cal P}}_{i-1}$, exactly as before. Once the Kalman state sequences ${\hat {\bf X}}_{i}$ and ${\hat {\bf X}}_{i+1}$ have been calculated, the system and output matrices can be recovered from an overdetermined system of linear equations, obtained from the stochastic state-space equations \cite{PeetersDeRoeck2000}.

\subsection{Multiple Output Backward Autoregressive (MOBAR)}\label{MOBAR_algorithm}

More classical system identification methods identify a dynamical model that do not contain the state variable. An example of those identification methods is the following particular ARMAV model (a multiple-input multiple-out ARMA model), which is equivalent to the (discrete) stochastic state-space model given by Eqs.\ (\ref{statespace1})-(\ref{statespace2}) 
\begin{equation} 
{\bf y}_{k} - {\bf A}_{1} \, {\bf y}_{k-1} - {\bf A}_{2} \, {\bf y}_{k-2} - \dots - {\bf A}_{p} \, {\bf y}_{k-p} = {\bf B}_{1} \, {\bf u}_{k-1} + {\bf B}_{2} \, {\bf u}_{k-2} + \dots + {\bf B}_{p} \, {\bf u}_{k-p} + {\boldsymbol{\epsilon}}_{k} 
\label{eq:difference}
\end{equation}
The matrices ${\bf A}_{i} \in \mathbb{R}^{\ell \times \ell}$ are the auto-regressive (AR) matrix parameters; the matrices ${\bf B}_{i} \in \mathbb{R}^{\ell \times u}$ are he moving average (MA) matrix coefficientes; ${\bf u}_{i} \in \mathbb{R}^{u \times 1}$ denotes the discrete multiple inputs to the system; and ${\boldsymbol{\epsilon}}_{k} \in \mathbb{R}^{\ell \times 1}$ denotes the white noise vector sequence. The MA part of the aforementioned ARMAV model represents the measurement noise and the input forces on the stochastic state-space model, whereas the AR part represents the system's physical properties \cite{OMABrincker}. Modal parameters are obtained by investigating the free response of the system; hence, the MA part of the ARMAV model is henceforth set to zero, leading to the following AR model
\begin{equation} 
{\bf y}_{k} - {\bf A}_{1} \, {\bf y}_{k-1} - {\bf A}_{2} \, {\bf y}_{k-2} - \dots - {\bf A}_{p} \, {\bf y}_{k-p} = {\bf 0}
\label{ARmodel}
\end{equation}
A way to solve the homogeneous Eq.\ (\ref{ARmodel}) in order to extract the modal parameters is to define the discrete state vector ${\bf z}_{k}$ by stacking $p$ response vectors ${\bf y}_{i}$, $i = \{k, k-1, \dots, k-p+1\}$, and a companion matrix ${\mathbb{A}}_{C}$, as follows
\begin{equation} 
{\bf z}_{k}=\begin{bmatrix}
{\bf y}_{k-p+1}\\
\vdots\\
{\bf y}_{k-1}\\
\vdots\\
{\bf y}_{k}\\
\end{bmatrix}
\;\;\;
{\mathbb{A}}_{C} = \begin{bmatrix}
{\bf 0} & {\bf I} & {\bf 0} & {\bf 0} \\
\vdots & {\bf 0} & \ddots & \vdots \\
{\bf 0} & \vdots & & {\bf I} \\
{\bf A}_{p} & {\bf A}_{p-1} & \cdots & {\bf A}_{1} \\
\end{bmatrix}
\label{augmentedvector_companionmatrix}
\end{equation}
where ${\bf I}\in \mathbb{R}^{\ell \times \ell}$ and ${\bf 0}\in \mathbb{R}^{\ell \times \ell}$ are, respectively, the identity and null matrices. Some manipulations lead to the evolution of the state vector ${\bf z}_{k}$ as follows: 

%ion For the next time step $k + 1$, the AR model of Eq.\ (\ref{ARmodel}) may be written  as 
%
%\begin{equation} 
%{\bf y}_{k+1} - {\bf A}_{1} \, {\bf y}_{k} - {\bf A}_{2} \, {\bf y}_{k-1} - \dots - {\bf A}_{p} \, {\bf y}_{k-p+1} = {\bf 0}.
\label{ARmodelnexttimestep}
%\end{equation}
%

%
\begin{equation} 
{\mathbb{A}}_{C} \, {\bf z}_{k} = {\bf z}_{k+1}
\label{MOBAR}
\end{equation}
One should notice that the companion matrix advances the discrete state vector one time step and one may look upon Eq.\ (\ref{ARmodelnexttimestep}) as a backward AR model with multiple outputs, hence the name MOBAR. The eigenvalues and eigenvectors of the companion matrix are the discrete eigenvalues and eigenvectors of the vibrating system, from which one computes the eigenfrequencies and damping ratios \cite{OMABrincker}. 

The AR matrix coefficients are computed with a procedure called Poly-Reference Time Domain \cite{vold1982multi}. The first step of the algorithm is to form a Hankel matrix ${\mathbb{H}}_{1} \in \mathbb{R}^{p \ell \times N-p}$ with $p$ block rows and $N-p$ columns containing the free responses at $p$ consecutive time instants and a single block row Hankel matrix ${\mathbb{ H}}_{2}\in \mathbb{R}^{\ell \times N-p}$.  The final step is to solve an over-determined system of linear equations given in Eq.\ (\ref{eq:AR2}) as follows 

\begin{equation} 
\label{eq:AR2}
{\mathbb{B}}_{C} \,  {\mathbb{H}}_{1} = {\mathbb{H}}_{2}
\end{equation}
\noindent where ${\mathbb{B}}_{C}$ denotes the last block row of the companion matrix ${\mathbb{A}}_{C}$ and detailed information about the components of matrices ${\mathbb{H}}_{1}$ and ${\mathbb{H}}_{2}$ are found in Appendix Eq.(\ref{eq:H1Ap}).

The main advantage of MOBAR over traditional AR models is that computational modes are automatically eliminated once stable modes are outside the unit circle of discrete time \cite{Tufts}.

\section{Proposed Strategy to Combine OMA Algorithms}\label{sec:UQ_OMA}

The algorithms presented in Sections \ref{SSI_algorithms} and \ref{MOBAR_algorithm} provide estimates of the undamped natural frequencies $\{ \hat{\omega}_1, \hat{\omega}_2, \ldots\, \hat{\omega}_{m}\}$, modal damping ratios $\{ \hat{\zeta}_1, \hat{\zeta}_2, \ldots, \hat{\zeta}_{m}\}$ and mode shapes $\{ \hat{\boldsymbol{\phi}}^{(1)}, \hat{\boldsymbol{\phi}}^{(2)} \ldots, \hat{\boldsymbol{\phi}}^{(m)}\}$ for each measured data set $\mathbf{y}^{(m)}(t)$, $t \in [0,T_a]$, where $T_a$ denotes the time interval used for analysis and $m = 0, 1, ..., M$ with $M$ denoting the number of available data sets.  One should notice that the estimators of the modal properties are random variables inasmuch as each set of measured data $\mathbf{y}^{(0)}(t), \mathbf{y}^{(1)}(t), \ldots, \mathbf{y}^{(M)}(t)$ may be thought as independent realizations of a stochastic process $\{\mathbf{Y}(t)\} \equiv \{\mathbf{y}(t;\upsilon) \in \mathbb{R}^{\ell}, t \in \mathbb{R}^{+}, \upsilon \in \Upsilon \}$, with $\Upsilon$ being the sample space. As a consequence, the estimates $\boldsymbol{\theta}_{k}^{(j)} = \{\hat{\omega}^{(j)}_k, \hat{\zeta}^{(j)}_k, \hat{\boldsymbol{\phi}}
^{(j)}_k\}^{T}$, $j = 0, 1, 2, \dots, M$, vary for each measured data set $\mathbf{y}^{(j)}(t)$; moreover, each OMA algorithm should be viewed as an independent estimator that provides its own set of estimates for the modal properties. 
Regarding information about the statistics of an estimator, it can be obtained as new realizations of the measured data become available.
\begin{figure}
    \centering
    \includegraphics[scale=0.6]{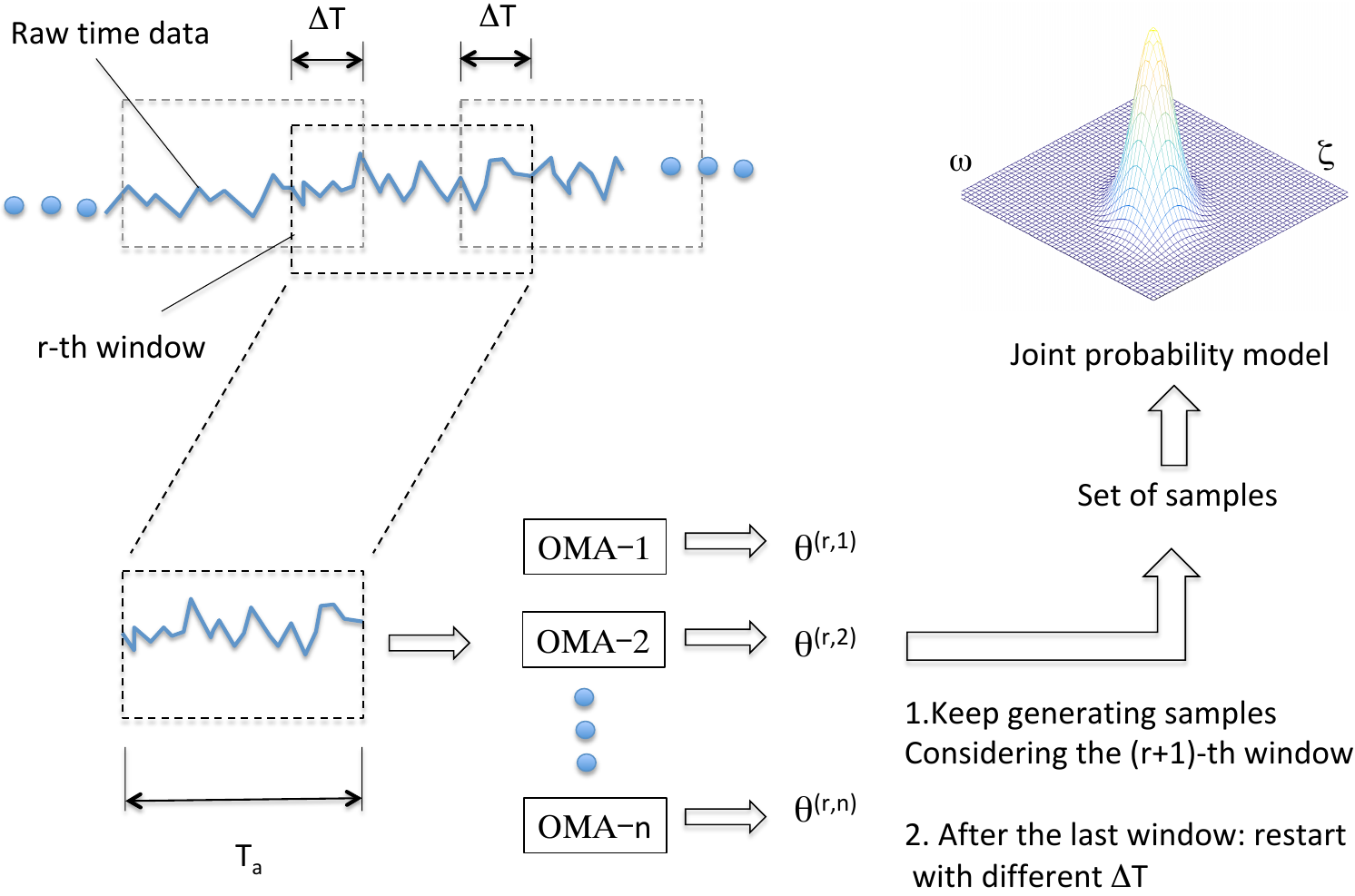}\\
    \caption{Proposed strategy: Raw time data is partitioned into time windows of length $T_a$ considering an overlapping $\Delta T$. Data from the $r$-th time window is provided as the input to the $j$-th OMA algorithm which in turn provides an estimate of modal parameters $\boldsymbol{\theta}^{(j,r)}$. The set of samples is enriched by restarting the whole process with different overlapping lengths $\Delta T$ to generate new samples. The whole set of samples is used to built a joint probability model.}
    \label{fig:Strategy}
\end{figure}
%

%$t \in \mathbb{R}^{+}$, $m = 0, 1, ..., M$ ($M$ denotes the number of available data sets).  One should notice that the estimators of the modal properties are random variables inasmuch as each set of measured data $\mathbf{y}^{(0)}(t), \mathbf{y}^{(1)}(t), \ldots, \mathbf{y}^{(M)}(t)$ may be thought as independent realizations of a stochastic process $\{\mathbf{Y}(t)\} \equiv \{\mathbf{y}(t;\upsilon) \in \mathbb{R}^{\ell}, t \in \mathbb{R}^{+}, \upsilon \in \Upsilon \}$, with $\Upsilon$ being the sample space. As a consequence, the estimates $\{\hat{\omega}_j, \hat{\zeta}_j, \hat{\boldsymbol{\phi}}
%^{(j)}\}$, $j = 1, 2, \dots, n$, vary for each measured data set; moreover, each OMA algorithm should be viewed as an estimator that provides its own set of estimates for the modal properties. 
%$\{ \hat{\omega}_1, \hat{\omega}_2, \ldots\}$, modal damping ratios $\{ \hat{\zeta}_1, \hat{\zeta}_2, \ldots, \}$ and mode shapes $\{ \hat{\boldsymbol{\phi}}^{(1)}, \hat{\boldsymbol{\phi}}^{(2)} \ldots, \}$. 
%Regarding information about the statistics of an estimator, it can be obtained as new realizations of the measured data become available. 
%In particular, when the user is dealing with numerical experiments, the statistics of the estimators may be obtained by means of Monte Carlo simulation analysis. 

From the practical point of view, the user will face a set of estimates provided by each one of the five OMA methods presented in Sections \ref{SSI_algorithms} and \ref{MOBAR_algorithm} and could eventually opt to select only the estimates provided by a subset  $\mathcal{S}_{\rm OMA} \subset \{ \mbox{DATA-SSI-UPC}, \mbox{DATA-SSI-CVA}, \mbox{DATA-SSI-PC}, \mbox{COV-SSI}, \mbox{MOBAR}\}$ of these algorithms, for example. In this context, as an effort to conjugate information provided by the OMA algorithms present in the set $\mathcal{S}_{\rm OMA}$ chosen for analyses, the authors propose to build mixed models that describe the characteristics of the modal properties estimates provided by all methods in $\mathcal{S}_{\rm OMA}$ simultaneously. 

The proposed approach is described by the illustration shown in Fig.\ref{fig:Strategy}. Firstly, time series raw data is partitioned into time windows of length $T_a$ considering an overlapping $\Delta T$ between adjacent windows. Data from the $r$-th time window is provided as the input to the $j$-th OMA algorithm which, in turn, provides an estimate of modal parameters $\boldsymbol{\theta}_{k}^{(j,r)}$. Secondly, aiming at enriching the set of samples, one restarts the whole process considering different overlapping lengths $\Delta T$ to generate new samples. The whole set of samples $\{ \{ \boldsymbol{\theta}_{k}^{(r,j)}\}\vert_{\Delta T_1}$, $\{ \boldsymbol{\theta}_{k}^{(r,j)}\}\vert_{\Delta T_2}, \ldots \}$  encompasses data from different OMA algorithms in $\mathcal{S}_{\rm OMA}$ and different overlapping lengths $\Delta T$. As each OMA algorithm possesses its own characteristics, the whole set of samples $\{ \{ \boldsymbol{\theta}_{k}^{(r,j)}\}\vert_{\Delta T_1}$, $\{ \boldsymbol{\theta}_{k}^{(r,j)}\}\vert_{\Delta T_2}, \ldots \}$ may possibly be composed of clusters in the parameter domain.

%Next we present the strategy adopted to build an approximate probability density function for $\boldsymbol{\theta}_k$.}

In the present work we assume that the modal parameters $\{\omega_k, \zeta_{k}\}^{T}$ of  the $k$-th mode of the system under analyses may be modelled as a random vector $\boldsymbol{\Theta}_k \in \mathbb{R}
^2$ whose densities $\pi(\boldsymbol{\Theta}_k)$ are multivariate Gaussian distributions, viz.
%Eq. Random Theta
\begin{equation}
\boldsymbol{\Theta}_k \sim \mathcal{N}(\boldsymbol{\mu}_k,\boldsymbol{\Sigma}_k)
\label{eq:Gaussian_models}
\end{equation}
\noindent where ${\boldsymbol{\mu}}_k$ and ${\boldsymbol{\Sigma}}_k$ correspond to the {\it{empirical}} mean and {\it{empirical}} covariance matrix computed with the  whole set of samples $\{ \{ \boldsymbol{\theta}^{(r,j)}\}\vert_{\Delta T_1}$, $\{ \boldsymbol{\theta}^{(r,j)}\}\vert_{\Delta T_2}, \ldots \}$ which enconpass all OMA methods in $\mathcal{S}_{\rm OMA}$  simultaneously. Although one understands that modelling the estimator for $\{\omega_j, \zeta_{j}\}^{T}$ as a Gaussian multivariate random vector is a simplifying hypothesis, it proved amenable for the current applications, as shortly demonstrated. After building the Gaussian probability models, one may use them to generate samples $\{\boldsymbol{\theta}_k
^{(1)}, \boldsymbol{\theta}_k
^{(2)}, \ldots, \boldsymbol{\theta}_k
^{(N'_{mc})}\}$ from the density $\mathcal{N}(\boldsymbol{\mu}_k,\boldsymbol{\Sigma}_k)$ of the modal properties that can be used for analyses, system identification, model validation, structural health monitoring and damage identification, for example. Further, with the Gaussian probability models, one may also draw $P_{\alpha}$ confidence ellipsis which is the contour of an elliptic region that encompasses the samples of the random variable $\boldsymbol{\Theta}_k \in \mathbb{R}^2$ with $\alpha\%$ probability.

\section{Actual Centrifugal Compressor}
\label{sec:shaft_description}

The dynamical system investigated is an actual centrifugal compressor that recycles a mixture of components made predominantly of hydrogen gas. The compressor is located at a Brazilian refinery's catalyst reformer unit (CRU) and its detailed information can be found in \cite{Hunter,zague2018operational}. This compressor was chosen due to the fact that it underwent through a stability verification testing (SVT) at its manufacturer's facility prior to shipment. Thus, experimental modal data extracted during stability tests exist for comparison with the estimates provided by the strategy proposed in the current work. Stability measurements mean here measures of damping ratio or, equivalently, of logarithmic decrement for a given set of modes. Interested readers are referred to  \cite{Hunter} for complete details about the entire shop SVT process, including modeling predictions and corrections.

Regarding damping estimates under design conditions, the logarithmic decrement of the first forward (1F) mode of the compressor, henceforth named  $\delta_{\rm{1F|SVT}}$, was estimated at the maximum continuous speed (MCS).  These estimates were provided by a simplified model correction approach (see the slope correction approach in Ref.\ \cite{Hunter}) based on the SVT measurements and they were in the range $\delta_{\rm {1F|SVT}
}
\in [0.12, 0.14]$. As for measurements during field operation, one expects to obtain estimates for the 1F mode stability within this predicted range, thus being similar to or slightly greater than the base stability measured at the shop.

Regarding sensors, the compression train is monitored by 16 displacement sensors, properly assembled to collect information from orthogonal directions named as $X$ and $Y$. Once this compressor train is equipped with flexible couplings, that isolates the lateral vibration from the motor, gearbox and compressor, only the two pairs of displacement sensors located at the compressor rotor were adopted for the purpose of modal parameters identification performed herein. One pair is located near the drive end (DE) journal bearing and the other near the non-drive end (NDE) journal bearing. 

\begin{figure}\
    \centering
	\includegraphics[scale=0.85, clip, trim=0cm 1cm 0cm 1cm]{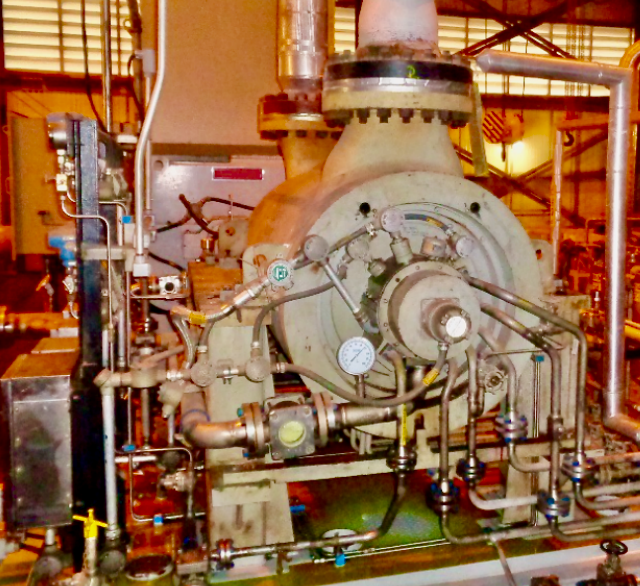}
	\caption{ Compressor that recycles a mixture of components made predominantly of hydrogen gas.}
	\label{fig:Photo}
\end{figure}

\begin{figure}\
    \centering
	\includegraphics[scale=0.45, clip, trim=0cm 1cm 0cm 1cm]{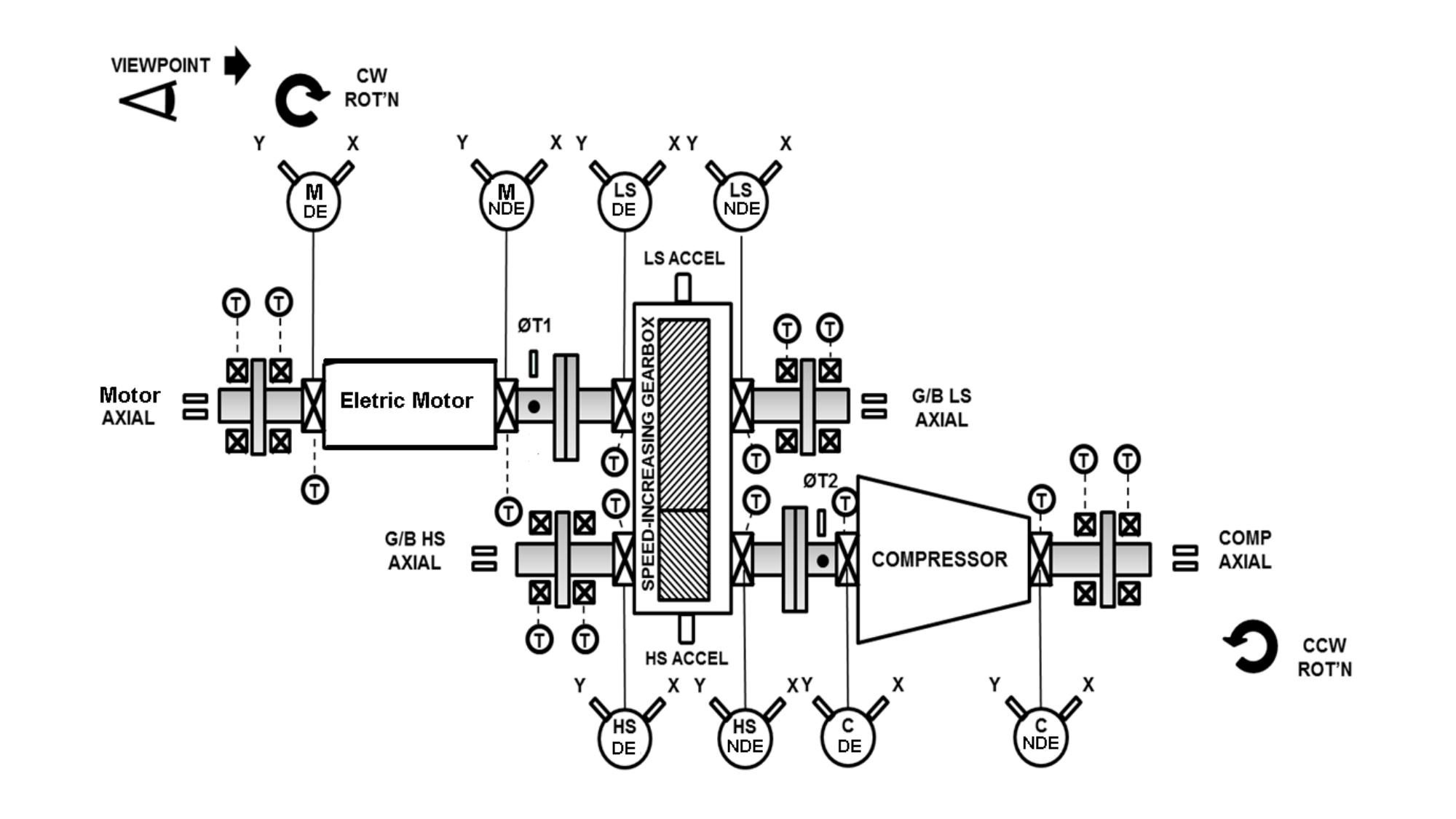}
	\caption{Illustrative sketch presenting the set of vibration and bearing temperature sensors of the compressor train. The symbol $T$ stands for temperature sensors, $X$, $Y$ stands for directions where displacements sensors monitoring radial displacement take place, $\phi$ stands for the phase measurements, $G/B$, $LS$, $HS$, $M$ and $C$ stands for gearbox, low speed shaft, high speed shaft, motor and compressor respectively. }
	\label{fig: P&ID}
\end{figure}

Figure \ref{fig:Photo} presents a photograph of the compressor and Fig.\ref{fig: P&ID} shows an illustrative diagram presenting the vibration and bearing pad temperature sensors of the complete compressor train, helping visualization of the four radial displacement sensors of the compressor. %As for the assessment of the stability of the compressor using OMA methods
Data from the four proximity sensors were collected during field operation. Data collection was performed by connecting a data collector to the Machinery Protection System buffered output channels. The sampling frequency was set to 128 kHz and the raw waveforms for each sensor were collected simultaneously during 25 minutes. The operational conditions during the process of data acquisition in the field are shown in Table \ref{tab:table1}, along with the Design and Shop Stability Test conditions. 

\begin{table}[h!]
	\begin{center}
		
		\caption{Design, Shop test and Field conditions. $P_s$ and $T_s$ correspond to the pressure and temperature at the suction. $P_d$ and $T_d$ correspond to the pressure and temperature at the discharge.  }
		\label{tab:table1}
		\begin{tabular}{l|c|c|c}
			\textbf{} & \textbf{Design} & \textbf{Shop} & \textbf{Field}\\ % <-- added & and content for each column
			%			$\alpha$ & $\beta$ & $\gamma$ & $\delta$ \\ % <--
			\hline
			Aero Case & Guarantee & Vacuum & H$_2$+HC\\ % <--
			Speed (rpm) & 13660 & 13660 & 13860\\ % <--
			$P_s$ (barA) & 9.03 & - & 9.74\\ % <--
			$T_s$ ($^{\circ}$C)& 38.0 & - & 31.6\\ % <--
			$P_d$ (barA) & 23.54 & - & 23.34\\ % <--
			$T_d$ ($^{\circ}$C) & 125.0 & - & 119.5\\ % <--
			Oil Inlet ($^{\circ}$C)& 43.3 - 54.4 & 46.6 & 42\\ % <--
			
		\end{tabular}
	\end{center}
\end{table}

\begin{figure}
	\centering
	% Requires \usepackage{graphicx}
	\includegraphics[scale=0.45,clip, trim=0cm 4.5cm 0cm 4.5cm]{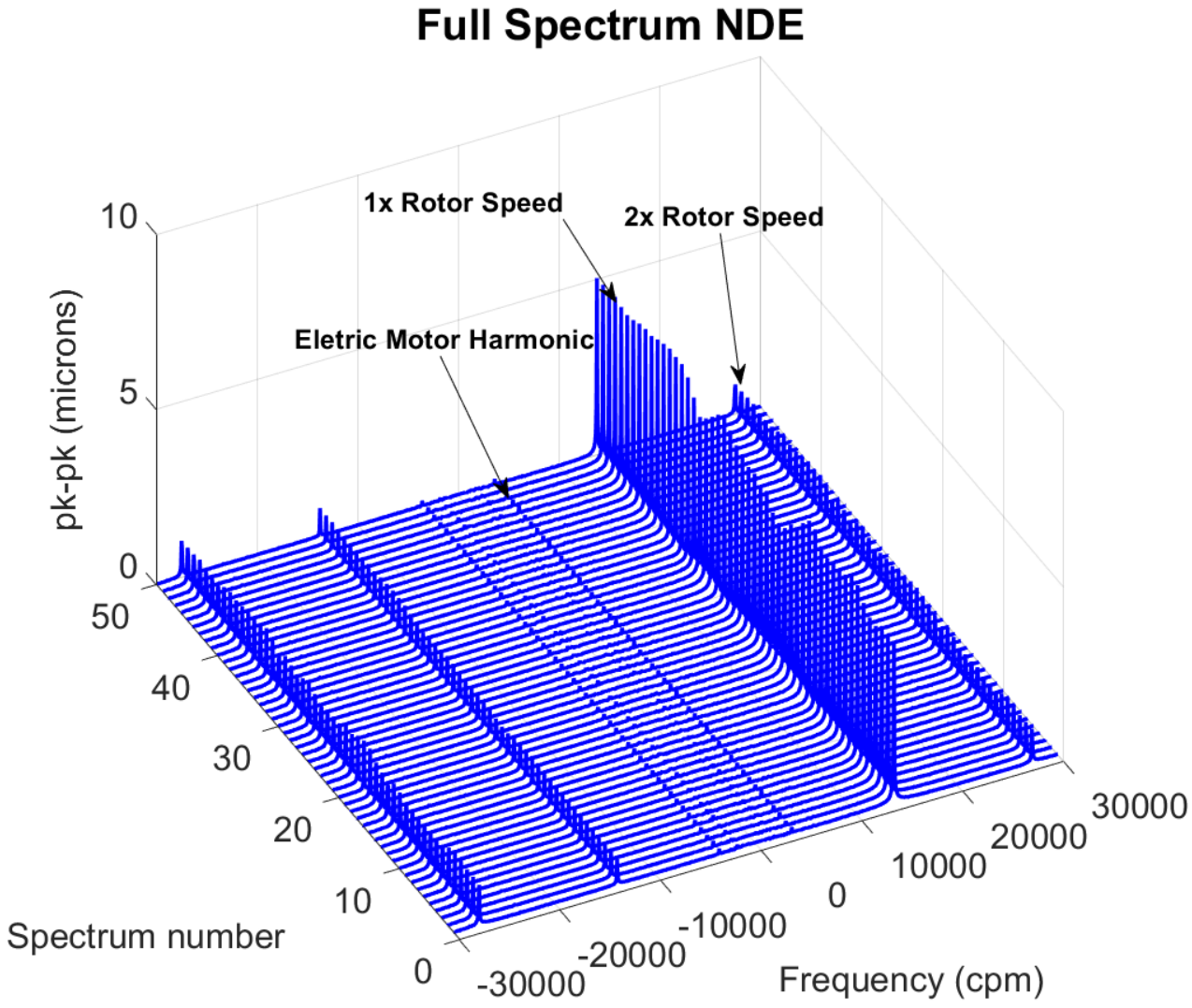}\\
	\caption{Full spectrum $\tilde{F}(\omega)$ computed with data $d_X(t)$ and $d_Y(t)$ measured at the Non Drive End during field operations at the maximum continuous speed (MCS). Spectrum number denotes the number of the realization for which the full spectrum was computed.}
	\label{fig: Field waterfall}
\end{figure}

\begin{figure}
	\centering
	% Requires \usepackage{graphicx}
	\includegraphics[scale=0.45,clip, trim=0cm 4.5cm 0cm 4.5cm]{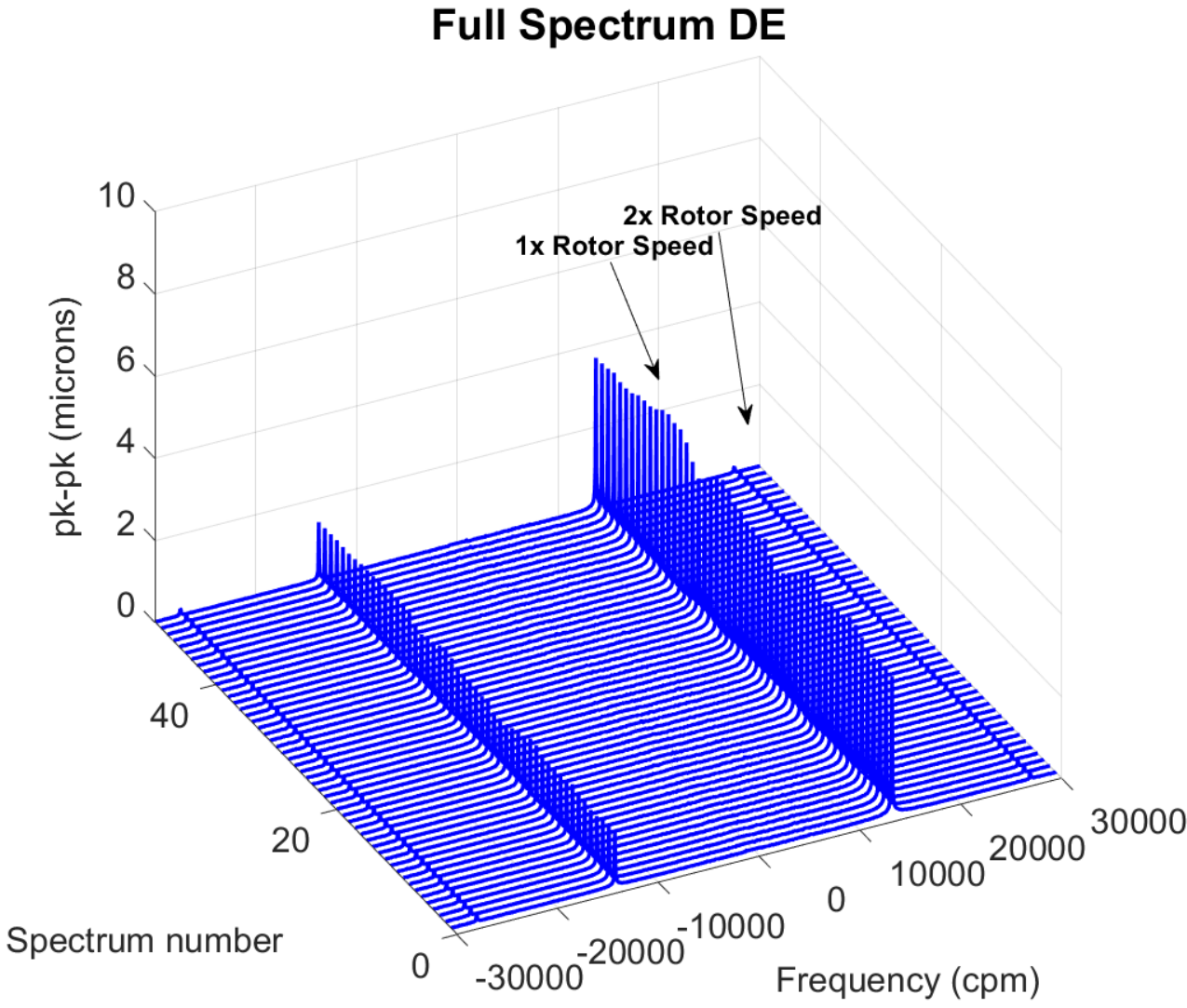}\\
	\caption{Full spectrum $\tilde{F}(\omega)$ computed with data $d_X(t)$ and $d_Y(t)$ measured at the Drive End during field operations at the maximum continuous speed (MCS). Spectrum number denotes the number of the realization for which the full spectrum was computed.}
	\label{fig: Field waterfall2}
\end{figure}

At its startup in 2012, the compressor was identified as being stable with no sign of subsynchronous vibrations in the proximity displacement probes spectra, except for the electric motor harmonic frequency. Aiming at providing means for the complete spectrum analyses of the orbits made by the axis of the compressor, the time domain displacement sensor measurements along $X$ and $Y$ directions are used to create a complex-valued feature $Z(t)$ defined as
\begin{equation}
Z(t) = d_X(t) + {\tt i} \, d_Y(t)
\end{equation}
\noindent where $d_X(t)$ and $d_Y(t)$ correspond to the time domain displacment measurements along $X$ and $Y$ directions and ${\tt i} = \sqrt{-1}$ stands for the imaginary unity. The full spectrum $\tilde{F}(\omega)$ is computed as the absolute value of the Fourier transform of the feature $Z(t)$, i.e., $\tilde{F}(\omega) = \vert \tilde{Z}(\omega) \vert$. The full spectrum computed with measured data from the Drive-End and Non Drive-End sensors are shown in Figs.\ \ref{fig: Field waterfall} and \ref{fig: Field waterfall2}, respectively. Finally, it should be highlighted that the absence of any significant subsynchronous vibration does not provide any quantitative measure of the first forward (1F) mode logarithmic decrement; moreover,  it could simply be the result of low ambient gas excitations being present in this hydrogen service, where the average gas mass density in the compressor is only 5.43 kg/m$^3$ \cite{zague2018operational}.


\section{Results and Discussions}

This section reports the modal parameter estimates provided by the proposed strategy using  
measured data recorded for the centrifugal compressor described in the previous section, when it was under operational conditions.

Measurement of time-domain response of the compressor at both the drive-end and non-drive-end were acquired for 25 minutes. On the other hand, the estimates of the modal properties $\{\omega_d, \zeta \}_m^{T}$ provided by the OMA algorithms are computed for time-domain signals with shorter duration (4 minutes), obtained by splitting the 25 minutes time-domain raw data signal into successive $T_a=$4 minutes signals and considering different overlapping lengths $\Delta T$ as schematically described in Fig.\ref{fig:Strategy}. Each splitted signal may be thought as an independent realization of the longer duration signal. %took into account measured data within time intervals of 4 minutes. Further, in order to create a scenario where one could make use of a set of realizations for the OMA algorithms, a strategy based on overlapping time intervals was taken into account. 
Hence, a set of measurement realizations was provided to OMA algorithms considering 4 minutes time intervals with $50\%$, $66.7\%$ and $75\%$ overlapping between two successive realizations. 

%Tables \ref{tab:table2} and \ref{tab:table3} report the empirical mean values of the damping ratios and damped natural frequencies for the first forward (1F) and first backward modes (1B). The empirical mean values were computed for all realizations (i.e, by considering $50\%$, $66.7\%$ and $75\%$ overlapping between two successive realizations). There are also reported in Tabs.\ \ref{tab:table2} and \ref{tab:table3} the corresponding modal parameters identified during factory stability test with magnetic exciters and for the compressor running in vacuum at the maximum continuous speed; these modal estimates are referred to as MOBAR$^{\square}}$. From Tabs.\ \ref{tab:table2} and \ref{tab:table3} one should notice that the variances for the 1B modal parameters provided by all OMA algorithms are much larger than those for the 1F mode, except for the variance in 1B modal damping ratio provided by DATA-SSI-CVA method, which is almost identical to the corresponding one for 1F mode. The higher spread on backward mode estimates may be attributed to two main factors: (i) low ambient excitation of first backward mode provided by internal flow; and (ii) higher damping, posing additional difficulties to identification, as earlier stated. Nevertheless, good agreement among estimates obtained with different OMA algorithms gives confidence in the results for a smooth running compressor, exposed to very low density gas excitation, with considerable scratches in the sensors' tracks.

\begin{figure}
\centering
\includegraphics[scale=0.75, clip, trim=3.5cm 7.5cm 3.5cm 8.5cm]{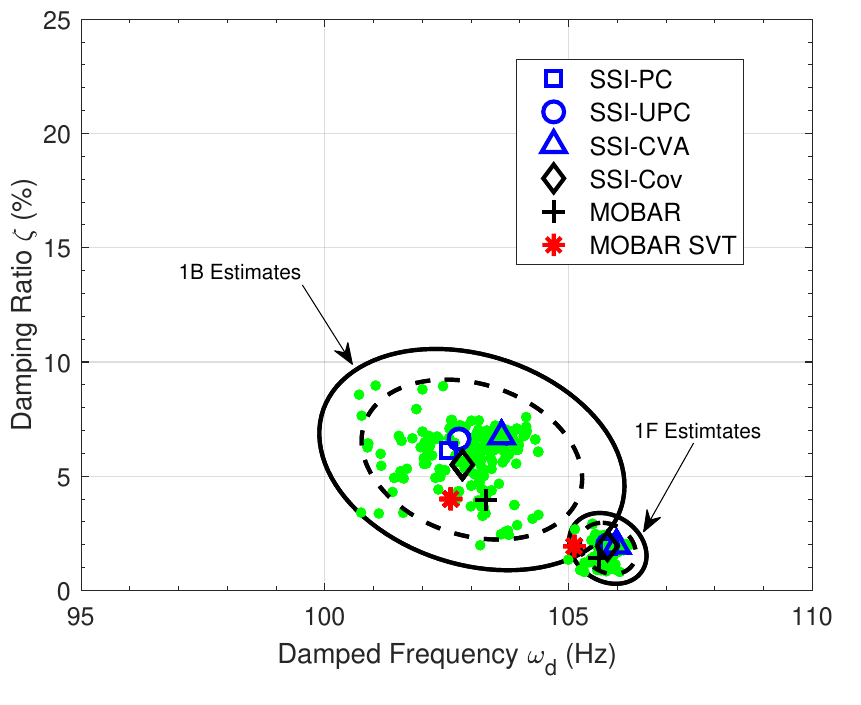}
\caption{Modal parameters $\{\omega_d, \zeta\}^{T}_m$ of the ﬁrst forward (1F) and first backward (1B) modes of the compressor. Symbols $\{ {\color{blue}{\square, \circ, \triangle, }} \Diamond, + \}$ denote the empirical mean values of the estimates provided by the OMA methods (only for reference). The ellipsis in dashed black line and the ellipsis in continuous black line describe the contour of the regions which encompass the samples of the Gaussian mixture model for the variable $\{\omega_d,\zeta \}^{T}_m$ with 95\% and 99.7\% probability, respectively. Note: green dots {\textcolor{green}{$\bullet$}} correspond to the samples provided by the mixture models and the red asterisk {\textcolor{red}{*}} correspond to the estimate provided by the standard  stability  verification  testing (SVT) of the centrifugal compressor prior to shipment.}
\label{fig: Hydrogen estimation}
\end{figure}

The proposed strategy is thus applied to combine the modal estimates provided by different OMA algorithms in order to build an approximate Gaussian joint probability distribution for the modal properties $\{\omega_d, \zeta\}^{T}_m$, as described in Section \ref{sec:UQ_OMA}. 
%The models shown in Eq.\ (\ref{eq:Gaussian_models}) were built using estimates provided by all OMA algorithms simultaneously. 
Table \ref{tab:table4} reports the components of the empirical mean ${\boldsymbol{\mu}}_m$ of the Gaussian probability models whereas the empirical covariance matrices for the first backward and forward modes, $\boldsymbol{\Sigma}_{\rm {1B}}$ and $\boldsymbol{\Sigma}_{\rm {1F}}$, are given in Eqs.\ (\ref{cov1B}) and (\ref{cov1F}). 
\begin{eqnarray}
{\boldsymbol{\Sigma}}_{\rm {1B}} =
\begin{bmatrix}
{ 0.8283} & { -0.3034} \\
{ -0.3034} & { 1.983}
\end{bmatrix}
\label{cov1B}
\end{eqnarray}
\begin{eqnarray}
{\boldsymbol{\Sigma}}_{\rm {1F}}
 =
\begin{bmatrix}
{ 0.0532} & { -0.0217} \\
{ -0.0217} & { 0.2031}
\end{bmatrix}
\label{cov1F}
\end{eqnarray}
Figure \ref{fig: Hydrogen estimation} shows the 95\% and 99.7\% ellipses provided by the corresponding Gaussian probability models for the modal parameters. Independent samples of the modal parameters $\{\omega_d, \zeta\}^{T}_{\rm 1B}$ and $\{\omega_d, \zeta\}^{T}_{\rm 1F}$ are drawn from the approximate Gaussian mixture models $\mathcal{N}(\boldsymbol{\mu}_{\rm 1B}, {\boldsymbol{\Sigma}}_{\rm {1B}})$ and $\mathcal{N}(\boldsymbol{\mu}_{\rm 1F}, {\boldsymbol{\Sigma}}_{\rm {1F}})$, respectively, and are also shown as green dots ${\color{green}{\bullet}}$ in Fig. \ref{fig: Hydrogen estimation}.   

From the analysis of both Tab.\ \ref{tab:table4} and Fig.\ \ref{fig: Hydrogen estimation} one may draw the following important conclusions. Firstly, uncertainties on the modal parameters $\{\omega_d, \zeta \}_{\rm 1B}^{T}$ of the first backward mode (1B) are much larger than those of the first forward mode (1F). Secondly, the first backward and forward damped natural frequency and modal damping ratio seem to be negatively correlated. Thirdly, the range predicted for the modal damping ratio during SVT ($\zeta_{\rm {1F|SVT}} \in [1.9, 2.2]$) is within the 99.7\% confidence region provided by the joint Gaussian probability distribution, as shown in Fig.\ \ref{fig: Hydrogen estimation} (the larger ellipses). 
%the mean value of the first forward mode logarithmic decrement, $\delta_{\rm 1F}$, is approximately 0.12, which agrees with the range predicted by the first simplified model correction approach from the shop SVT ($\delta_{\rm{1F|SVT}}\in [0.12, 0.14]$).
Fourthly, one finds a good agreement among the mean values of the modal parameters estimated with different OMA algorithms (indicated as symbols in Fig.\ \ref{fig: Hydrogen estimation}) and those obtained from shop SVT (red asterisk symbol). As expected, there are slight differences among these mean values, which are higher for the first backward mode. Nevertheless, these mean values are all within the cluster of samples drawn from the joint Gaussian probability distribution (indicated as green dots in Fig.\ \ref{fig: Hydrogen estimation}). Finally, the proposed strategy provides not only a single value for the modal parameters but also their joint probability distributions, from which important joint and marginal statistics are readily obtained (e.g., marginal variances and confidence regions). Such uncertainty quantification of the modal parameters, even though approximate, takes into account estimates provided by different OMA algorithms, is computationally efficient and thus has practical appealing for online structural health monitoring goals. 

\begin{table}[h!]
	\begin{center}
		\caption{Mean value and coefficient of variation (inside square brackets) obtained for the first forward and backward modal properties of the compressor. Note: The column with header MOBAR$^{\square}$ reports the modal parameters obtained with MOBAR algorithm using as input data the measured data collected during the standard verification testing (SVT) of the compressor.}
		\label{tab:table4}
		\begin{tabular}{l|c|c|c|c|c|c|}
			\textbf{} &  \textbf{Gaussian Mixture Model} & \textbf{MOBAR}$^{\square}$ \\ % <-- added & and content for each column
			%			$\alpha$ & $\beta$ & $\gamma$ & $\delta$ \\ % <--
			\hline
			Operational Condition &Field & Vacuum \\ \hline % <--
			1st Forward $\omega_d$(Hz) &  105.8 [0.0022] & 105.12 \\ % <--
			1st Forward $\zeta (\%)$ & 1.84 [0.244] & 1.94  \\ % <--
			1st Backward $\omega_d$(Hz) & 103.01 [0.0088] & 102.58  \\ % <--
			1st Backward $\zeta (\%)$ &5.73 [0.245] & 4.01 \\ % <--

		\end{tabular}
	\end{center}
\end{table}

\section{Final Remarks}

%The main goal of the current work was to perform a critical assessment of three widely-used time-domain OMA methods when applied to time-domain responses computed or measured for rotating machines, not only by comparing the estimates obtained for the modal parameters (eigenfrequencies, damping ratios and mode shapes) but also the associated uncertainties, induced by simulated or actual variations in vibration test configurations.

The main goal of the current work was to propose a computationally-efficient and robust statistical procedure to combine estimates of the modal parameters provided by any set of OMA algorithms so as to avoid preference for a particular OMA algorithm (and thus biasing the end results) and also to derive an approximate Gaussian joint probability distribution of the modal parameters, from which marginal and joint statistics of interest such as mean value, variance and confidence regions are readily provided. The proposed statistical strategy avoided full Bayesian computation, but also takes into account measurement uncertainties as well as model uncertainties inherent to each OMA algorithm. The proposed approach was applied to long-duration time-domain field data collected for an actual centrifugal compressor. The estimates obtained for both forward and backward modal parameters were compared against those obtained during standard stability verification testing (SVT) of the investigated centrifugal compressor, prior to shipment, using classical EMA algorithms. 

Although the OMA algorithms employed in the current work are well-acknowledged by the scientific research community, their application to identification of modal parameters of centrifugal compressors is far from being completely understood by the rotating machinery community. To the authors' point of view, the current work also contributes to the literature on rotating machinery by providing a statistical strategy able to combine estimates obtained with different OMA algorithms in order to perform an uncertainty quantification of the identified modal parameters. The results reported for the dynamical system investigated revealed that the proposed strategy provides fast and accurate estimates for both the modal parameters and the associated uncertainties, which may improve not only future rotordynamic designs but also the predictive accuracy of existing computational models and real-time assessment of rotor's stability condition (by inspecting the magnitude of the real part of the identified eigenvalues).

We finish by remarking that fast and accurate estimates for the modal parameters of actual centrifugal compressors are thus feasible by employing the proposed statistical strategy with well-known OMA algorithms. The approximate uncertainty quantification approach proposed here may handle large amounts of data commonly available for rotating machinery in a computationally-efficient manner. Although much less rigorous than previous Bayesian OMA approaches discussed in the literature, the proposed approach does not need to rely on simplifying hypotheses such as well-separated natural frequencies and small damping in order to improve its computational performance. The aforementioned hypotheses are almost always satisfied by large-scale civil engineering structures, but not always verified for large-scale rotating machines as industrial centrifugal compressors. 

\section{Acknowledgments}

The authors would like to thank Leonardo Ishimoto from Petrobras and Doctors. José A. Vazquez and C. Hunter Cloud from BRG, who provided valuable support during this investigation. Daniel A. Castello would like to acknowledge  \textit{Coordenação de Aperfeiçoamento de Pessoal de Nível Superior} (CAPES) - Finance code 001 - Grant PROEX 803/2018, and the Brazilian agency: \textit{Conselho Nacional de Desenvolvimento Científico e Tecnológico} (CNPQ) - Grant number 312355/2020-3. Carlos F.\ T.\ Matt would like to acknowledge the Electric Energy Research Center (CEPEL) for the financial support of the research described in the current manuscript.

%%%%%%%%%%%%%%%%%%%%%%%
    
%%%%%%%%%%%%%%%%%%%%%%%
\bibliographystyle{elsarticle-num}     
\bibliography{Bibliography}

\begin{thebibliography}{10}
\expandafter\ifx\csname url\endcsname\relax
  \def\url#1{\texttt{#1}}\fi
\expandafter\ifx\csname urlprefix\endcsname\relax\def\urlprefix{URL }\fi
\expandafter\ifx\csname href\endcsname\relax
  \def\href#1#2{#2} \def\path#1{#1}\fi

\bibitem{Sedehietal2020}
\protect{Sedehi, O., Katafygiotis, L.\ S.\ and Papadimitriou, C.}, Hierarchical
  bayesian operational modal analysis: Theory and computations, Mechanical
  Systems and Signal Processing 140 (2020) 1--27.

\bibitem{Auetal2018}
\protect{Au, S.-K., Brownjohn, J.\ M.\ W.\ and Mottershead, J.\ E.},
  Quantifying and managing uncertainty in operational modal analysis,
  Mechanical Systems and Signal Processing 102 (2018) 139--157.

\bibitem{Salehietal2018}
\protect{Salehi, M., Esfarjani, S.\ M.\ and Ghorbani, M.}, Modal parameter
  extraction of a huge four stage centrifugal compressor using operational
  modal analysis method, Latin American Journal of Solids and Structures 15~(3)
  (2018) 1--11.

\bibitem{Wagneretal2022}
\protect{Wagner, G., Lima, R.\ and Sampaio, R.}, Modal identification of a
  light and flexible wind turbine blade under wind excitation, Journal of
  Engineering Mathematics 133~(3) (2022).
\newblock \href {https://doi.org/10.1007/s10665-022-10210-1}
  {\path{doi:10.1007/s10665-022-10210-1}}.

\bibitem{JVC2}
Z.-R. Lu, D.~Yang, L.~Huang, L.~Wang, Covariance regression for operational
  modal analysis, Journal of Vibration and Control 28~(11-12) (2022)
  1295--1310.

\bibitem{OMABrincker}
\protect{Brincker, R.\ and Ventura, C.}, Introduction to Operational Modal
  Analysis, 2nd Edition, Wiley, Chichester, United Kingdom, 2015.

\bibitem{reynders2012system}
\protect{Reynders, E.}, System identification methods for (operational) modal
  analysis: review and comparison, Archives of Computational Methods in
  Engineering 19~(1) (2012) 51--124.

\bibitem{akaike1974markovian}
\protect{Akaike, H.}, Markovian representation of stochastic processes and its
  application to the analysis of autoregressive moving average processes,
  Annals of the Institute of Statistical Mathematics 26~(1) (1974) 363--387.

\bibitem{bendat1980engineering}
\protect{Bendat, J.\ S.\ and Piersol, A.\ G.}, Engineering applications of
  correlation and spectral analysis, New York, Wiley-Interscience, 1980. 315 p.
  (1980).

\bibitem{ibraham1977method}
\protect{Ibraham, S.\ R.}, A method for the direct identification of vibration
  parameter from the free responses, Shock and Vibration Bulletin 47~(4)
  (1977).

\bibitem{vold1982multi}
\protect{Vold, H., Kundrat, J., Rocklin, G.\ T.\ and Russell, R.}, A
  multi-input modal estimation algorithm for mini-computers, SAE Transactions
  (1982) 815--821.

\bibitem{Pappa}
\protect{Juang, J.\ and Pappa, R.}, \protect{An eigensystem realization
  algorithm (ERA) for modal parameter identification and model reduction},
  Journal of Guidance, Control and Dynamics 8~(5) (1985) 620--627.

\bibitem{NExT}
\protect{James, G.\ H., Carne, T.\ G.\ and Lauffer, J.\ P.}, The natural
  excitation technique (next) for modal parameter extraction from operating
  structures, International Journal of Analytical and Experimental Modal
  Analysis 10~(4) (1995) 260.

\bibitem{van2012subspace}
\protect{Van Overschee, P.\ and De Moor, B.\ L.}, Subspace identification for
  linear systems: Theory—Implementation—Applications, Springer Science \&
  Business Media, 1996.

\bibitem{yuen2001bayesian}
\protect{Yuen, K.-V.\ and Katafygiotis, L.\ S.}, Bayesian time--domain approach
  for modal updating using ambient data, Probabilistic Engineering Mechanics
  16~(3) (2001) 219--231.

\bibitem{YuenKatafygiotis2003}
\protect{Yuen, K.-V.\ and Katafygiotis, L.\ S.}, Bayesian fast fourier
  transform approach for modal updating using ambient data, Advances in
  Structural Engineering 6~(2) (2003) 81--95.

\bibitem{yuen2011bayesian}
\protect{Yuen, K.-V.\ and Kuok, S.-C.}, Bayesian methods for updating dynamic
  models, Applied Mechanics Reviews 64~(1) (2011) 010802.

\bibitem{reynders2008uncertainty}
\protect{Reynders, E.\, Pintelon, R.\ and De Roeck, G.}, Uncertainty bounds on
  modal parameters obtained from stochastic subspace identification, Mechanical
  Systems and Signal Processing 22~(4) (2008) 948--969.

\bibitem{modak2010harmonics}
\protect{Modak, S.\ V., Rawal, C.\ and Kundra, T.\ K.}, Harmonics elimination
  algorithm for operational modal analysis using random decrement technique,
  Mechanical Systems and Signal Processing 24~(4) (2010) 922--944.

\bibitem{peeters2007removing}
\protect{Peeters, B., Cornelis, B., Janssens, K.\ and Van der Auweraer, H.},
  Removing disturbing harmonics in operational modal analysis, in: Proceedings
  of International Operational Modal Analysis Conference, Copenhagen, Denmark,
  2007.

\bibitem{qi2008vibration}
K.~Qi, Z.~He, Z.~Li, Y.~Zi, X.~Chen, Vibration based operational modal analysis
  of rotor systems, Measurement 41~(7) (2008) 810--816.

\bibitem{gres2019orthogonal}
\protect{Gres, S., Andersen, P., Hoen, C.\ and Damkilde, L.}, Orthogonal
  projection-based harmonic signal removal for operational modal analysis, in:
  Structural Health Monitoring, Photogrammetry \& DIC, Volume 6, Springer,
  2019, pp. 9--21.

\bibitem{reynders2016uncertainty}
E.~Reynders, K.~Maes, G.~Lombaert, G.~De~Roeck, Uncertainty quantification in
  operational modal analysis with stochastic subspace identification:
  validation and applications, Mechanical Systems and Signal Processing 66
  (2016) 13--30.

\bibitem{pintelon2007uncertainty}
\protect{Pintelon, R., Guillaume, P.\ and Schoukens, J.}, Uncertainty
  calculation in (operational) modal analysis, Mechanical Systems and Signal
  Processing 21~(6) (2007) 2359--2373.

\bibitem{Guglielmo2}
\protect{Guglielmo, A., Mitaritonna, N., Catanzaro, M.\ and Libraschi, M.},
  \protect{Full load stability test (FLST) on LNG compressor}, in: ASME Turbo
  Expo 2014: Turbine Technical Conference and Exposition, American Society of
  Mechanical Engineers, 2014, pp. V07AT31A006--V07AT31A006.

\bibitem{Guglielmo15}
\protect{Guglielmo, A., Baldassarre, L., Catanzaro, M., Zague, L.\ O.,
  Ishimoto, L., Miranda, M.\ A.\ and Silva, R.\ T.}, Operational modal analysis
  application for the measure of logarithm decrement in centrifugal compressor,
  in: Proceedings of the 44th Turbomachinery Symposium, Turbomachinery
  Laboratories, Texas A\&M Engineering Experiment Station, 2015.

\bibitem{Loyds}
\protect{Carden, E.\ P., Sehlstedt, N., Nielsen, K.\ K., Lundholm, S.\ and
  Morosi, S.}, Stability analysis and assessment of rotor trains using
  operational modal analysis, in: Proceedings of the 9th IFToMM International
  Conference on Rotor Dynamics, Springer, 2015, pp. 2083--2093.

\bibitem{Cloud2007}
\protect{Cloud, H.}, Stability of rotors supported by tilting pad journal
  bearings, Ph.d. thesis, University of Virginia, Charlottesville, Va, USA
  (2007).

\bibitem{Hunter_damp}
\protect{Cloud, C. H., Maslen, E.\ H.\ and Barrett, L.\ E.}, Damping ratio
  estimation techniques for rotordynamic stability measurements, Journal of
  Engineering for Gas Turbines and Power 131~(1) (2009) 012504.

\bibitem{Holopainen}
\protect{Holopainen, T. P., Aatola, S. A., Cloud, C.\ H.\ and Li, G.},
  Comparison of modal parameter estimation techniques for electromechanical
  rotordynamics of cage induction motors, in: ASME 2007 International Design
  Engineering Technical Conferences and Computers and Information in
  Engineering Conference, American Society of Mechanical Engineers, 2007, pp.
  1215--1224.

\bibitem{Hunter}
\protect{Pettinato, B.\ C., Cloud, C.\ H.\ and Campos, R.\ S.}, Shop acceptance
  testing of compressor rotordynamic stability and theoretical correlation, in:
  Proceedings of the 39th Turbomachinery Symposium, Texas A\&M University.
  Turbomachinery Laboratories, 2010.

\bibitem{Noronha}
\protect{Noronha, R.\ F., Miranda, M.\ A., Lucchesi, K.,Memmott, E.\ A.\ and
  Ramesh, K.}, \protect{Stability Testing of CO2 Compressors}, in: Proceedings
  of the 43rd Turbomachinery Symposium, Texas A\&M University. Turbomachinery
  Laboratories, 2014.

\bibitem{JVC1}
S.~Chauhan, Using the unified matrix polynomial approach (umpa) for the
  development of the stochastic subspace identification (ssi) algorithm,
  Journal of Vibration and Control 19~(13) (2013) 1950--1961.

\bibitem{PeetersDeRoeck1999}
\protect{Peeters, B.\ and De Roeck, G.}, Reference based stochastic subspace
  identification for output-only modal analysis, Mechanical Systems and Signal
  Processing 13~(6) (1999) 855--878.

\bibitem{Andersen}
\protect{Brincker, R.\ and Andersen, P.}, \protect{Understanding Stochastic
  Subspace Identification}, in: Proceedings of the 24th IMAC, St. Louis,
  Missouri, 2006, pp. 279--311.

\bibitem{PeetersDeRoeck2000}
\protect{Peeters, B.\ and De Roeck, G.}, Reference based stochastic subspace
  identification in civil engineering, Inverse Problems in Engineering 8~(1)
  (2000) 47--74.

\bibitem{Tufts}
\protect{Kumaresan, R.\ and Tufts, D.}, Estimating the parameters of
  exponentially damped sinusoids and pole-zero modeling in noise, IEEE
  Transactions on Acoustics, Speech, and Signal Processing 30~(6) (1982)
  833--840.

\bibitem{zague2018operational}
\protect{Zague, L.\ O., Cloud, C.\ H., Miranda, M.\ A.\ and Campos, R.\ S.},
  Operational modal analysis of a centrifugal compressor in field operation,
  in: International Conference on Rotor Dynamics, Springer, 2018, pp. 460--474.

\end{thebibliography}
	
	\newpage

	%%%%%%%%%%
	\appendix
	\section{{\bf{Appendix - System Matrices}}}\label{Appendix}
	
	Some matrices introduced in Section \ref{sec:oma_algorithms} are detailed in this Appendix.\\
	
		    \begin{equation} \label{eq:Hankel1Ap}
{\mathbb{H}} = \frac{1}{\sqrt{N}}
\begin{bmatrix}
{\bf y}_{0} & {\bf y}_{1} & {\bf y}_{2} & \dots & {\bf y}_{N-1} \\
{\bf y}_{1} & {\bf y}_{2} & {\bf y}_{3} & \dots & {\bf y}_{N} \\
{\bf y}_{2} & {\bf y}_{3} & {\bf y}_{4} & \dots & {\bf y}_{N+1} \\
\vdots & \vdots & \vdots & \ddots & \vdots \\
{\bf y}_{i-1} & {\bf y}_{i} & {\bf y}_{i+1} & \dots & {\bf y}_{i+N-2} \\ \cline{1-5}
{\bf y}_{i} & {\bf y}_{i+1} & {\bf y}_{i+2} & \dots & {\bf y}_{i+N-1} \\
{\bf y}_{i+1} & {\bf y}_{i+2} & {\bf y}_{i+3} & \dots & {\bf y}_{i+N} \\
{\bf y}_{i+2} & {\bf y}_{i+3} & {\bf y}_{i+4} & \dots & {\bf y}_{i+N+1} \\
\vdots & \vdots & \vdots & \ddots & \vdots \\
{\bf y}_{2i-1} & {\bf y}_{2i} & {\bf y}_{2i+1} & \dots & {\bf y}_{2i+N-2}
\end{bmatrix} = 
\begin{bmatrix}
{\bf Y}_{0|i-1} \\ \cline{1-1}
{\bf Y}_{i|2i-1}
\end{bmatrix} =
\begin{bmatrix}
{\bf Y}_{p} \\ \cline{1-1}
{\bf Y}_{f}
\end{bmatrix}
\end{equation}
%%%
\begin{equation} \label{eq:Toeptitz1Ap}
{\bf T}_{1|i} = {\bf Y}_{f} \, {\bf Y}_{p}^{T} = 
\begin{bmatrix}
{\bf R}_{{\bf y},i} & {\bf R}_{{\bf y},i-1} & {\bf R}_{{\bf y},i-2} & \dots & {\bf R}_{{\bf y},1} \\
{\bf R}_{{\bf y},i+1} & {\bf R}_{{\bf y},i} & {\bf R}_{{\bf y},i-1} & \dots & {\bf R}_{{\bf y},2} \\
{\bf R}_{{\bf y},i+2} & {\bf R}_{{\bf y},i+1} & {\bf R}_{{\bf y},i} & \dots & {\bf R}_{{\bf y},3} \\
\vdots & \vdots & \vdots & \ddots & \vdots \\
{\bf R}_{{\bf y},2i-1} & {\bf R}_{{\bf y},2i-2} & {\bf R}_{{\bf y},2i-3} & \dots & {\bf R}_{{\bf y},i} \\
\end{bmatrix}
\end{equation}
Regarding the output covariance matrix at any time lag $i$ for discrete output signals ${\bf y}_{k}$, $k = 0, 1, \dots, N-1$, one finds 
\begin{equation}
{\bf R}_{{\bf y},i} \equiv {\mathbb{E}}[{\bf y}_{k+i} \, {\bf y}_{k}^{T}] = \lim_{N \rightarrow \infty} \sum_{k=0}^{N-1} \frac{1}{\sqrt{N}} \, {\bf y}_{k+i} \, \frac{1}{\sqrt{N}} \, {\bf y}_{k}^{T} = \lim_{N \rightarrow \infty} \frac{1}{N} \sum_{k=0}^{N-1} {\bf y}_{k+i} \, {\bf y}_{k}^{T}
\label{covarianceestimateAp}
\end{equation}
%%%
\begin{equation} %\label{eq:Toeplitz2Ap}
{\bf T}_{1|i}=
\begin{bmatrix}
{\bf C} \, {\bf A}^{i-1} \, {\bf G} & {\bf C} \, {\bf A}^{i-2} \, {\bf G} & {\bf C} \, {\bf A}^{i-3} \, {\bf G} & \dots & {\bf C} \, {\bf G} \\
{\bf C} \, {\bf A}^{i} \, {\bf G} & {\bf C} \, {\bf A}^{i-1} \, {\bf G} & {\bf C} \, {\bf A}^{i-2} \, {\bf G} & \dots & {\bf C} \, {\bf A} \, {\bf G} \\
{\bf C} \, {\bf A}^{i+1} \, {\bf G} & {\bf C} \, {\bf A}^{i} \, {\bf G} & {\bf C} \, {\bf A}^{i-1} \, {\bf G} & \dots & {\bf C} \, {\bf A}^{2} \, {\bf G} \\
\vdots & \vdots & \vdots & \ddots & \vdots \\
{\bf C} \, {\bf A}^{2i-2} \, {\bf G} & {\bf C} \, {\bf A}^{2i-3} \, {\bf G} & {\bf C} \, {\bf A}^{2i-4} \, {\bf G} & \dots & {\bf C} \, {\bf A}^{i-1} \, {\bf G} \\
\end{bmatrix} = {\boldsymbol \Gamma}_{i} \, {\boldsymbol \Delta}_{i}
\label{blockHankelmatrixdecompAp}
\end{equation}

%%%

\begin{eqnarray} 
{\bf T}_{2|i+1} &=&
\begin{bmatrix}
{\bf R}_{{\bf y},i+1} & {\bf R}_{{\bf y},i} & {\bf R}_{{\bf y},i-1} & \dots & {\bf R}_{{\bf y},2} \\
{\bf R}_{{\bf y},i+2} & {\bf R}_{{\bf y},i+1} & {\bf R}_{{\bf y},i} & \dots & {\bf R}_{{\bf y},3} \\
{\bf R}_{{\bf y},i+3} & {\bf R}_{{\bf y},i+2} & {\bf R}_{{\bf y},i+1} & \dots & {\bf R}_{{\bf y},4} \\
\vdots & \vdots & \vdots & \ddots & \vdots \\
{\bf R}_{{\bf y},2i} & {\bf R}_{{\bf y},2i-1} & {\bf R}_{{\bf y},2i-2} & \dots & {\bf R}_{{\bf y},i+1} \\
\end{bmatrix} \nonumber \\
&=& \begin{bmatrix}
{\bf C} \, {\bf A}^{i} \, {\bf G} & {\bf C} \, {\bf A}^{i-1} \, {\bf G} & {\bf C} \, {\bf A}^{i-2} \, {\bf G} & \dots & {\bf C} \, {\bf A} \, {\bf G} \\
{\bf C} \, {\bf A}^{i+1} \, {\bf G} & {\bf C} \, {\bf A}^{i} \, {\bf G} & {\bf C} \, {\bf A}^{i-1} \, {\bf G} & \dots & {\bf C} \, {\bf A}^{2} \, {\bf G} \\
{\bf C} \, {\bf A}^{i+2} \, {\bf G} & {\bf C} \, {\bf A}^{i+1} \, {\bf G} & {\bf C} \, {\bf A}^{i} \, {\bf G} & \dots & {\bf C} \, {\bf A}^{3} \, {\bf G} \\
\vdots & \vdots & \vdots & \ddots & \vdots \\
{\bf C} \, {\bf A}^{2i-1} \, {\bf G} & {\bf C} \, {\bf A}^{2i-2} \, {\bf G} & {\bf C} \, {\bf A}^{2i-3} \, {\bf G} & \dots & {\bf C} \, {\bf A}^{i} \, {\bf G} \\ 
\end{bmatrix} = {\boldsymbol \Gamma}_{i} \, {\bf A} \, {\boldsymbol \Delta}_{i}
\label{blockHankelmatrix2decompAp}
\end{eqnarray}
%%%
\begin{equation} 
\label{eq:H1Ap}
{\mathbb{H}}_{1} = \begin{bmatrix}
{\bf y}_{1} & {\bf y}_{2} & \cdots & {\bf y}_{N-p}\\
{\bf y}_{2} & {\bf y}_{3} & \cdots & {\bf y}_{N-p+1} \\
\vdots & \vdots & \ddots & \vdots \\
{\bf y}_{p} & {\bf y}_{p+1} &  & {\bf y}_{N-1}\\
\end{bmatrix} 
\;\;\;\;
{\mathbb{H}}_{2}=\begin{bmatrix}
{\bf y}_{p+1} & {\bf y}_{p+2} & \cdots & {\bf y}_{N} \\
\end{bmatrix}
\end{equation}
%%%%%%%%%%
\end{document}